\documentclass[a4paper,9pt]{article}
\usepackage{graphicx}  
\usepackage{amsmath}   
\usepackage[compress]{cite}
\usepackage{amssymb} 
\usepackage{calligra}
\usepackage{bm} 
\usepackage{dcolumn}
\usepackage{color}
\usepackage{tensor}
\usepackage[T1]{fontenc}
\usepackage{colortbl}	
\usepackage{marvosym}
\usepackage[dvipsnames]{xcolor}
\usepackage{tensor}
\usepackage{textcomp}
\usepackage{mathrsfs}
\usepackage{gensymb}
\usepackage{appendix}
\usepackage{float}
\usepackage{amsfonts}
\usepackage{varioref}
\usepackage{physics}
\usepackage{booktabs}
\usepackage{textcomp}
\usepackage{enumerate}
\usepackage{tikz}
\usepackage{multicol}
\usepackage{soul}
\usepackage{notoccite}
\usepackage{graphicx}
\usepackage{cite}
\usepackage{lineno}
\usepackage{palatino,mathpazo}
\usepackage[caption=true]{subfig}
\usepackage{parskip}
\RequirePackage{parskip}
\RequirePackage[colorlinks,citecolor=Cyan,urlcolor=Blue,linkcolor=Blue]{hyperref}
\allowdisplaybreaks

\DeclareMathAlphabet{\mathbold}{OML}{cmbrm}{b}{it}

\def\g{\mathbf{g}}
\labelformat{section}{Section #1} 
\labelformat{subsection}{Section #1} 
\labelformat{subsubsection}{Section #1}
\labelformat{subsubsubsection}{Section #1}
\labelformat{equation}{Eq.~(#1)} 
\labelformat{figure}{Fig.~[#1]} 
\labelformat{subfigure}{Fig.~[\thefigure#1]}
\labelformat{table}{Tab.~#1} 
\labelformat{appendix}{Appendix #1}
\definecolor{lime}{HTML}{A6CE39}
\DeclareRobustCommand{\orcidicon}{
	\begin{tikzpicture}
		\draw[lime, fill=lime] (0,0) 
		circle [radius=0.16] 
		node[white] {{\fontfamily{qag}\selectfont \tiny ID}};
		\draw[white, fill=white] (-0.0625,0.095) 
		circle [radius=0.007];
	\end{tikzpicture}
	\hspace{-2mm}
}
\foreach \x in {A, ..., Z}{\expandafter\xdef\csname orcid\x\endcsname{\noexpand\href{https://orcid.org/\csname orcidauthor\x\endcsname}
		{\noexpand\orcidicon}}
}

\newcommand{\authorA}{
  Subhadip Sau\thanks{\href{mailto:subhadipsau2@gmail.com}{subhadipsau2@gmail.com}}\orcidC$^{1,2}$}

\newcommand{\authorB}{John W. Moffat
\thanks{\href{mailto:jmoffat@perimeterinstitute.ca}{jmoffat@perimeterinstitute.ca}}~$^{3,4}$}
\title{ \textsc{Regular Rotating Black Hole:} \\\textsc{Probing the boundaries of Radiative Signatures and Jet Power }}

\author{\authorA, and  \authorB  \\
    \\
    \small{$^{1}$Department of Physics, Jhargram Raj College, Jhargram, West Bengal-721507, India} \\
    \small{$^{2}$Institute of Astronomy, Space and Earth Science (IASES), Bidhan Sishu Sarani, Kolkata-700054, India} \\
    \small{$^{3}$Perimeter Institute for Theoretical Physics, Waterloo, Ontario N2L 2Y5, Canada}\\
    \small{$^{4}$Department of Physics and Astronomy, University of Waterloo, Waterloo, Ontario N2L 3G1, Canada}
}
	
\begin{document}
\maketitle

\begin{abstract}
We perform a detailed observational analysis of several galactic X-ray binaries, focusing on the interplay between black hole spin, jet power, and radiative efficiency within the context of Blandford–Znajek-powered jets. Using updated measurements from continuum fitting and Fe-line methods, we constrain the spin parameter a and the deviation parameter $\beta$ for five key black hole systems: H1743-322, XTE J1550-564, GRS 1124-683, GRO J1655-40, and GRS 1915+105. For each system, we compare the allowed parameter spaces derived independently from observed radiative efficiencies and emitted jet powers under different assumptions for the jet Lorentz factor  $\Gamma=2,5$. By overlapping these observational constraints with theoretical expectations for regular black holes, we assess the viability of various spin-deviation combinations in explaining the observed phenomena. Our results reveal significant restrictions on the allowed values of $\beta$, with typical upper bounds around 0.38–0.4, except for rapidly spinning sources where the constraint becomes notably tighter. We further present a modified method for generating rotating solutions from static regular black hole spacetimes and provide a robust theoretical framework for relating jet power to black hole angular frequency in curved geometries. We also find that the theoretical jet power is modified by regularization factor for regular black holes. These findings place stringent observational bounds on deviations from the Kerr geometry and provide important insight into the astrophysical mechanisms powering accreting stellar-mass black holes.
\end{abstract}


\newpage
\section{Introduction}

The emergence of black holes as real astrophysical entities has been firmly established through multiple channels. Observations of gravitational waves from binary mergers\cite{LIGOScientific:2016aoc,LIGOScientific:2021qlt,PhysRevD.102.043015,Abbott_2020,PhysRevD.103.122002,PhysRevD.93.122003} and the imaging of black hole shadows\cite{Fish:2016jil,EventHorizonTelescope:2019dse,EventHorizonTelescope:2019uob,EventHorizonTelescope:2019ggy,EventHorizonTelescope:2022wkp,EventHorizonTelescope:2022urf} have not only confirmed their existence but also emphasized the need to understand the physics of extreme gravity near their horizons. Although the classical Kerr geometry of general relativity has proven successful in modeling rotating black holes, it carries a severe theoretical shortcoming. At the centre of a Kerr black hole lies a curvature singularity\cite{GEROCH1968526,PhysRev.187.1784}, a point where the gravitational field becomes infinitely strong and the known laws of physics cease to be meaningful.
To address this problem, several attempts have been made to construct black hole solutions that remain regular throughout the entire spacetime. Such regular black holes do not possess any singular region. Instead, they are supported either by modified matter fields\cite{PhysRevLett.80.5056,PhysRevLett.96.031103,PhysRevLett.96.251101} or by quantum-inspired corrections\cite{PhysRevD.76.104030,PhysRevD.82.104035,Fernandes:2023vux,PhysRevD.106.L021901,Nicolini:2019irw,PhysRevD.62.043008,PhysRevLett.132.031401} to the gravitational field equations. A compelling class of solutions arises within the framework of scalar-tensor-vector gravity (STVG), known more commonly as MOG or MOdified Gravity. This theory modifies Einstein’s gravity by introducing an additional vector field and scalar functions that effectively strengthen the gravitational interaction at different scales. Without invoking dark matter, MOG has been able to provide consistent explanations for galactic rotation curves\cite{Moffat:2013sja,Moffat:2014pia}, the dynamics of galaxy clusters\cite{Moffat:2013uaa,Brownstein:2007sr}, shadow structure of SGR A* and M87*\cite{Sau:2022afl}, and large-scale cosmological structures\cite{Davari:2021mge,Moffat:2014bfa,Moffat:2007ju,Moffat:2011rp,Moffat:2021log}.
Regular black holes in MOG theory represent an important development. They combine the resolution of central singularities with a framework that seeks to explain cosmic phenomena through modifications of gravity alone\cite{Moffat:1806.01903}. These spacetimes typically feature a nonsingular core where curvature invariants remain finite, along with a modified gravitational potential that affects geodesic motion near the black hole. While the theoretical structure of these solutions has been investigated, their astrophysical signatures remain relatively unexplored.

One important observational quantity that is highly sensitive to the near-horizon geometry is the radiative efficiency of an accreting black hole. This efficiency describes the fraction of rest mass energy that can be radiated away as matter spirals into the black hole. It depends primarily on the radius of the innermost stable circular orbit, which in turn is shaped by the details of the background geometry\cite{PhysRevD.87.023007}. In rotating spacetimes, this quantity also reflects the frame-dragging effects and the potential well experienced by infalling matter\cite{Bardeen1972RotatingBH}. Any deviation from the Kerr geometry, whether caused by regularization or by a different theory of gravity, may leave a measurable imprint on the radiative efficiency\cite{Banerjee:2020ubc,Narzilloev:2023btg,Cheng:2024lth}.

Energetic transient jets are frequently detected in astrophysical systems with black holes, and these jets are largely considered to be fuelled by the spinning energy of the black hole via magnetic fields, whose field lines are anchored in the horizon of  the black hole  or the accretion disc. Extensive review is given in \cite{10.1111/j.1365-2966.2004.07598.x,10.1093/mnras/198.2.345}.
Several high-precision observations have suggested that the efficiency of some black hole systems could be quite large. These include active galactic nuclei and black hole X-ray binaries where the disk emission can be tracked reliably. In this context, it is worthwhile to ask whether regular MOG black holes can accommodate such high efficiencies, or whether they predict a suppression or enhancement relative to the Kerr benchmark. Moreover, since the near-horizon geometry is also responsible for launching relativistic jets, the regular structure may influence not only the efficiency of radiation but also the total energetics of these systems.

In the present work, we investigate the radiative efficiency of regular rotating black holes that arise within the MOG theory. We analyze the structure of the innermost stable circular orbits and compute the efficiency for a range of spin parameters and MOG coupling values. Our goal is to assess the observational viability of these black holes and to determine whether they predict signatures that can be tested against current and future astrophysical data.

The structure of this paper is as follows.  \ref{SEC:MOG_Theory} presents a concise overview of the Scalar-Tensor-Vector theory of gravity, commonly known as MOdified Gravity (MOG), with particular emphasis on the range of parameters that remain theoretically viable within the framework. In  \ref{SEC:Framework}, we outline the theoretical approach employed to examine the radiative efficiency and jet power associated with black holes in this theory.  \ref{SEC:OBS} focuses on the observational constraints inferred from known microquasar systems and their implications for the MOG model. Lastly, in Section \ref{SEC:CON}, we provide a synthesis of our findings and offer a discussion on the broader significance of the results.

 We work with the mostly plus signature i.e. the spacetime signature $(-,+,+,+)$ is used. We choose the units such that speed of light $c=1$, Newton's constant $G_{\rm N}=1$. 

\section{Compact objects in STVG/MOG theory}\label{SEC:MOG_Theory}

\subsection{Field Equations  from STVG action}

The action associated with the Scalar-Tensor-Vector gravity theory is given by \cite{Moffat:2005si, Moffat:1806.01903, Moffat:1412.5424, Moffat:1603.05225, LopezArmengol:1611.05721, Perez:1705.02713, LopezArmengol:1611.09918}
\begin{flalign}\label{Eq:S1}
	\mathcal{S}=\mathcal{S}_{\rm GR}+\mathcal{S}_{\phi}+\mathcal{S}_{S}+\mathcal{S}_{\rm Matter}
\end{flalign}
where the terms are given by
\begin{subequations}
\begin{flalign}
\mathcal{S}_{\rm GR}&=\dfrac{1}{16}\int \dd ^{4}x\sqrt{-g} \dfrac{1}{G}R\\
\mathcal{S}_{\phi}&=-\int \dd ^{4}x\sqrt{-g}\left(\dfrac{1}{4}\mathcal{B}^{\mu\nu}\mathcal{B}_{\mu\nu}-\dfrac{1}{2}\mu^{2}\phi_{\mu}\phi^{\mu}-\mathcal{J}^{\mu}\phi_{\mu}\right)\\
\mathcal{S}_{S}&=\int\dd^{4}x\sqrt{-g}\dfrac{1}{G^{3}}\left[\dfrac{1}{2}g^{\alpha\rho}\nabla_{\alpha}G\nabla_{\rho}G -V(G)-\mathcal{J}G \right]+\int\dd^{4}x\sqrt{-g}\dfrac{1}{\mu^{2}G}\left[\dfrac{1}{2}g^{\alpha\rho}\nabla_{\alpha}\mu\nabla_{\rho}\mu-V(\mu) \right]
\end{flalign}
\end{subequations}
where $g_{\alpha\rho}$ is the spacetime metric, $R$ is the \emph{Ricci scalar}, $g$ is the determinant of the metric. Also, here $\phi^{\mu}$ represents the proca-like massive vector field 
and the corresponding strength tensor is given by
\begin{flalign} \mathcal{B}_{\mu\nu}\equiv\partial_{\mu}\phi_{\nu}-\partial_{\nu}\phi_{\mu}
\end{flalign}
It should be noted that $G(x)$ and $\mu(x)$ are the scalar fields of the theory with the corresponding potentials $V(G)$ and $V(\mu)$, respectively. In Eq.(1) $\mathcal{S}_{\rm Matter}$ represents the field action associated with the matter present in the spacetime. The energy-momentum tensor associated with the gravitational source can be written as
\begin{flalign}
T_{\mu\nu}=T_{\mu\nu}^{\phi}+T^{S}_{\mu\nu}+T_{\mu\nu}^{\rm Matter}
\end{flalign}
where
\begin{flalign}
	T_{\mu\nu}^{i}=-\dfrac{2}{\sqrt{-g}}\pdv{S_{i}}{g^{\mu\nu}}\hspace{2cm}i={\phi,S,{\rm Matter}}
\end{flalign}
Here $T_{\mu\nu}^{\phi},~T_{\mu\nu}^{S},~T_{\mu\nu}^{\rm Matter}$ represents the energy-momentum tensor associated with vector field $\phi^{\mu}$, scalar fields $G(x)$ and $\mu(x)$, and matter field, respectively. Also, $\mathcal{J}^{\mu}$ and $\mathcal{J}$ denotes the vector and scalar field currents, respectively.

Since we are interested in the galactic compact objects, we have to find the Schwarzschild-like or Kerr-like solution in STVG framework. To do so, we need to follow some  assumptions: (a) Matter energy-momentum tensor  along with the vector and scalar currents are taken to be zero. (b) Effect of the mass $\mu$ of the scalar field $\phi_{\mu}$ is prominent at kiloparsec scale from the source. So while solving the field equations, we assume the mass of the vector field is zero. (c) The constant $G$ depends on the characteristic parameter $\beta=\alpha\left( 1+\alpha\right)^{-1}$ by $G=G_{N}(1+\alpha)=\dfrac{G_{N}}{1-\beta}$, where $G_{N}$ is the Newton's gravitational constant and we assume that $\partial_{\mu}G=0$. Also the range of the dimensionless characteristic parameter is $\beta\in[0,1]$. Under these assumptions, from \ref{Eq:S1} we obtain
\begin{flalign}
\mathcal{S}=\dfrac{1}{16\pi G}\int\dd^{4}x\sqrt{-g}\left(R-\dfrac{1}{4}\mathcal{B}^{\mu\nu}\mathcal{B}_{\mu\nu} \right)
\end{flalign}	
Variation of the action given in Eq.(6) with respect to the spacetime metric $\g_{\mu\nu}$, we obtain
\begin{flalign}
G_{\mu\nu}\equiv R_{\mu\nu}-\dfrac{1}{2}g_{\mu\nu}R=8\pi G~ T_{\mu\nu}^{\phi}
\end{flalign}
where $T_{\mu\nu}^{\phi}$ is the energy-momentum tensor associated with the vector field $\phi_{\mu}$
is given by
\begin{flalign}
T_{\mu\nu}^{\phi}=\dfrac{1}{4\pi}\left(\tensor[]{\mathcal{B}}{_{\mu}^{\rho}}\mathcal{B}_{\mu\nu}-\dfrac{1}{4}g_{\mu\nu}\mathcal{B}^{\alpha\beta}\mathcal{B}_{\alpha\beta} \right)	
\end{flalign}	
In order to find the dynamical equation for the vector field, one need to vary Eq.(6) with respect to the vector field $\phi_{\mu}$, which eventually leads to
\begin{flalign}
\nabla_{\nu}\mathcal{B}^{\mu\nu}\equiv\dfrac{1}{\sqrt{-g}}\partial_{\nu}\left(\sqrt{-g}B^{\mu\nu} \right)=0
\end{flalign}
Morever, we should also note that the gravitational charge $\mathcal{Q}_{\rm grav}$ associated with the STVG/MOG vector field is proportional to the mass of the gravitational source as $\mathcal{Q}_{\rm grav}=\sqrt{\alpha G_{N}}M=\sqrt{\beta(1-\beta)G_{N}}M_{\beta}$, where $M_{\beta}=(1+\alpha)M$. This gravitational charge $\mathcal{Q}_{\rm grav}$ has a noticable effect on Newtonian acceleration in weak field limit and the acceleration of a particle is this case can be given by
\begin{flalign}
a(r)=-\dfrac{G_{N}M}{r^{2}}\left[1+\alpha-\alpha (1+\mu r)e^{-\mu r} \right]
\end{flalign}
For resonably small object, in weak field regime $\mu r\ll 1$, one can recover the Newtonian results. However, considering the parameter-post-Newtonian corrections, it can be shown that  STVG/MOG is consistent with the solar system experiments\cite{Moffat:2014asa}.

\subsection{Static, spherically symmetric, regular MOG spacetime}

The static, spherically symmetric, regular MOG spacetime solution is given by~\cite{Moffat:1806.01903}
\begin{equation}\label{Eq:S_11}
\dd s^{2} = -F(r)\dd t^{2} + \dfrac{1}{F(r)}\dd r^{2} + r^{2}\left( \dd \theta^{2} + \sin^{2}\theta \dd \phi^{2} \right)
\end{equation}
where
\begin{flalign}
	F(r)=1-\dfrac{2(1+\alpha)Mr^{2}}{\bigg[ r^{2}+\alpha(1+\alpha)M^{2}\bigg]^{3/2}}+\dfrac{\alpha(1+\alpha)M^{2}r^{2}}{\bigg(r^{2}+\alpha(1+\alpha)M^{2}\bigg)^{2}}
\end{flalign}
Here, $M$ is the mass parameter of the central gravitating object. the associated gravi-electric field is given by
\begin{flalign}
E_{\rm grav}(r)=\sqrt{\alpha}Mr^{4}\left[\dfrac{r^{2}-5\alpha(1+\alpha)M^{2}}{\bigg(r^{2}+\alpha(1+\alpha)M^{2}\bigg)}+\dfrac{15}{2}\dfrac{(1+\alpha)M}{\bigg(r^{2}+\alpha(1+\alpha)M^{2}\bigg)^{7/2}} \right]
\end{flalign}
For convenience, we introduce the new dimensionless parameter as
\begin{flalign}
\beta=\dfrac{\alpha}{1+\alpha}
\end{flalign}
The Arnowitt-Deser-Misner (ADM) mass of the central gravitating object is given by
\begin{flalign}
	M_{\rm ADM}=(1+\alpha)M=\dfrac{M}{1-\beta}\equiv M_{\beta}
\end{flalign} 
We can express the static, spherically symmetric, regular MOG spacetime by \ref{Eq:S_11} with
\begin{flalign}\label{Eq:RF}
F(r)=1-\dfrac{2M_{\beta}r^{2}}{\left(r^{2}+\beta M_{\beta}^{2} \right)^{3/2}}+\dfrac{\beta M_{\beta}^{2}r^{2}}{\left(r^{2}+\beta M_{\beta}^{2} \right)^{2}}
\end{flalign}
The gravitational source charge in terms of this new parameter is given by
\begin{flalign}
\mathcal{Q}_{\rm grav}=\sqrt{\beta(1-\beta)G_{N}}M_{\beta}
\end{flalign}
The structure of the horizon has been discussed in literature\cite{Sau:2022afl}. The critical value of the characteristic parameter $\beta_{\rm c}$ is deterimed by solving the following equation
\begin{flalign}\label{Eq:Crit}
10800\beta_{\rm c}^{3}-12096\beta_{\rm c}^{2}+20304\beta_{\rm c}-6912=0
\end{flalign}
\ref{Eq:Crit} has only one real solution and other two solutions are complex and conjugate to each other. The solution(real) of the equation yields
\begin{flalign}
\beta_{c}=\dfrac{1}{75}\left[28-\dfrac{2741}{\left(8902+8025\sqrt{321}\right)^{1/3} }+ \left(8902+8025\sqrt{321}\right)^{1/3}\right]\simeq0.402186
\end{flalign}

  At this critical value, we have extremal black hole, i.e, the black hole with single horizon. For $\beta<\beta_{\rm c}$, we have black holes with two horizons and for $\beta>\beta_{\rm c}$, we have horizonless compact objects\cite{Sau:2022afl}.

  \subsection{Regular MOG Rotating Compact Objects}
  
  The Newman--Janis algorithm is widely known for generating rotating black hole solutions starting from static, spherically symmetric ones. In our case, the static solution presented in \ref{Eq:S_11}, along with the redshift function defined in \ref{Eq:RF}, serves as the starting point. By employing a suitably modified version of the Newman--Janis procedure that ensures compatibility with the associated energy--momentum tensor, a consistent rotating solution within the framework of MOG can be derived~\cite{Azreg-Ainou:2014pra}(see \ref{App:NJA}).
  
  The resulting geometry describes a regular rotating black hole in MOG theory, expressed in Boyer--Lindquist coordinates as
  \begin{flalign}
  	\dd s^{2}=-\left[1-\dfrac{2 f(r)}{\Sigma(r,\theta)} \right]\dd t^{2} +\dfrac{\Sigma(r,\theta)}{\Delta(r)}\dd r^{2}+\Sigma(r,\theta) \dd\theta^{2}-\dfrac{4af(r)}{\Sigma(r,\theta)}\sin^{2}\theta  \dd t \dd \phi+ \dfrac{\mathcal{A}(r,\theta)}{\Sigma(r,\theta)} \sin^{2}\theta~ \dd \phi^{2}\label{Eq:Metric_kerr}
  \end{flalign}
  where the  functions associated with the metric are defined as
  \begin{subequations}
  	\begin{flalign}
  		\Sigma(r,\theta)&=r^{2}+a^{2}\cos^{2}\theta\\
  		f(r)&=\dfrac{1}{2}r^{2}\left(1-F(r) \right)\\
  		\Delta(r)&=r^{2}F(r)+a^{2}=r^{2}+a^{2}-2f(r)\\
  		\mathcal{A}(r,\theta)&=\left(r^{2}+a^{2} \right)^{2}-a^{2}\Delta(r)\sin^{2}\theta
  	\end{flalign}
  \end{subequations}
  Here, the parameter $a$ represents the spin of the black hole. The deviation from general relativity due to the MOG parameter $\beta$ is encoded in the function $F(r)$, which is given by
  \begin{flalign}
  	F(r)=1-\dfrac{2M_{\beta}r^{2}}{\left(r^{2}+\beta M_{\beta}^{2}\right)^{3/2}}+\dfrac{\beta M_{\beta}^{2}r^{2}}{\left(r^{2}+\beta M_{\beta}^{2} \right)^{2}}
  \end{flalign}
  
\begin{figure}[h!]
  	\centering
  	\includegraphics[width=9cm]{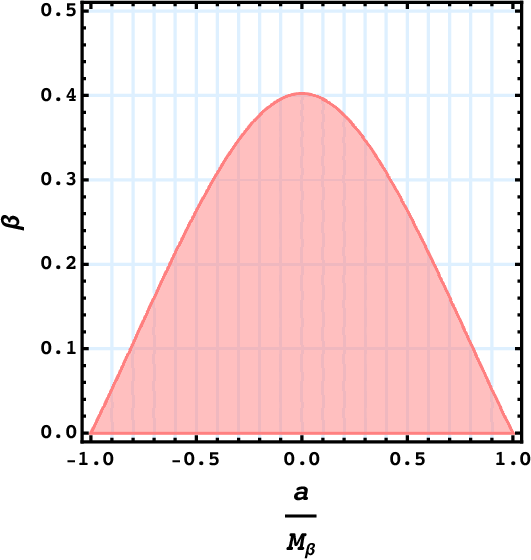}
  	\caption{The parameter space for the existence of horizon has been shown in the red-shaded zone. It indicates that the non-zero value of MOG parameter $\beta$ effectively reduces the permissible range of the spin value of a black hole. It also shows that the maximum allowed parameter is $\beta=\beta_{c}\simeq 0.4022$.}\label{Fig:PS}
  \end{figure}

  The parameter space for which the black hole possesses an event horizon is shown in \ref{Fig:PS}. The shaded region indicates the allowed values of spin $a$ and MOG parameter $\beta$. As $\beta$ increases, the range of admissible spin values shrinks. Beyond a critical threshold $\beta_c \simeq 0.4022$, no event horizon forms.
  
  To analyze particle motion, we confine it to the equatorial plane by setting $\theta = \pi/2$ and $\dot{\theta} = 0$. This simplification retains the essential physics while reducing the complexity of calculations. The equatorial slice of the metric then becomes
  \begin{flalign}
  	\dd s^{2}_{\rm Eq}=-F(r)\dd t^{2}+\dfrac{r^{2}}{\bigg[r^{2}F(r)+a^{2}\bigg]}\dd r^{2}-2a\left(1-F(r)\right)\dd t \dd \phi+\bigg[r^{2}+a^{2}+a^{2}\{1-F(r)\}\bigg]\dd \phi^{2}
  \end{flalign}
  
\begin{figure}[htbp!]
  	\centering
  	\includegraphics[scale=1]{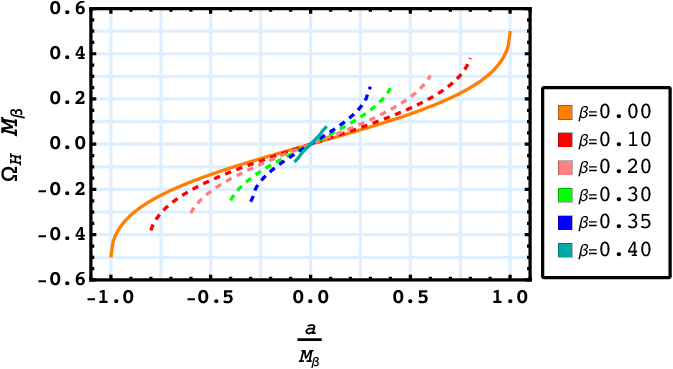}
  	\caption{The variation of angular velocity of horizon $\Omega_{H}$ with the change in the spin value $a$ has been shown for different values of MOG parameter $\beta$.}\label{Fig:AV}.
  \end{figure}

  An important physical quantity in the study of black hole jets is the angular velocity of the event horizon, denoted by $\Omega_H$. It plays a central role in energy extraction processes such as the Blandford--Znajek mechanism. For the metric under consideration, $\Omega_H$ is given by
  \begin{flalign}
  	\Omega_{H}=-\left.\dfrac{g_{t\phi}}{g_{\phi\phi}}\right\vert_{r=r_{H}}
  \end{flalign}
  The dependence of $\Omega_H$ on the spin $a$ and the MOG parameter $\beta$ is depicted in \ref{Fig:AV}. The influence of $\beta$ is to higher the horizon angular velocity for a given spin, thereby affecting the power and dynamics of jet emission in astrophysical scenarios.
  
  In their foundational work, Blandford and Znajek~\cite{1977MNRAS.179..433B} investigated the extraction of energy and momentum from a Kerr black hole immersed in a stationary, axisymmetric, force-free, magnetized plasma. Their analysis was carried out using Boyer--Lindquist coordinates. However, a known limitation of this coordinate system is that the metric component $g_{rr}$ diverges at the event horizon. This divergence introduces a coordinate singularity in the Maxwell equations, thereby requiring a careful imposition of boundary conditions at the horizon to ensure physical consistency of the electromagnetic field.
  
  To overcome this complication, we adopt the framework developed by McKinney and Gammie~\cite{McKinney:2004ka}, who employed Kerr--Schild coordinates. These coordinates are regular across the horizon, eliminating the need for additional boundary prescriptions at that surface. Extending this idea to a general axisymmetric black hole spacetime, we introduce ingoing Eddington--Finkelstein-like coordinates to carry out our analysis in a manner that remains regular at the horizon,
  \begin{subequations}
  	\begin{flalign}
  		\dd t'&=\dd t+\dfrac{\Sigma}{\Delta}\left[1-f(r,\theta) \right]\dd r\\
  		\dd \phi'&=\dd \phi +\dfrac{a}{\Delta}\dd r
  	\end{flalign}
  \end{subequations}
  The metric given in Eq.~\ref{Eq:Metric_kerr} can be expressed in \emph{Kerr--Schild} coordinates $(t',r,\theta,\phi')$ as
  \begin{flalign}\label{Eq:GKS}
  	\dd s^{2}&=-\left[1-\dfrac{2f(r)}{\Sigma(r,\theta)}\right]\dd t'^{2}+\left[1+\dfrac{2f(r)}{\Sigma(r,\theta)} \right]\dd r^{2}+\dfrac{4f(r,\theta)}{\Sigma(r,\theta)}\dd t' \dd r+\Sigma(r,\theta)\dd \theta^{2}\nonumber\\
  	&~~~~~+\left[\dfrac{(r^{2}+a^{2})^{2}-\Delta a^{2}\sin^{2}\theta}{\Sigma(r,\theta)} \right]\sin^{2}\theta\dd\phi'^{2}
  	-2\left[1+\dfrac{2f(r)}{\Sigma(r,\theta)} \right]a\sin^{2}\theta\dd r~\dd\phi'-\dfrac{4f(r)}{\Sigma(r,\theta)}a\sin^{2}\theta \dd t'\dd \phi'
  \end{flalign}
  The Kerr--Schild coordinate system proves essential for our subsequent analysis and is discussed further in \ref{App_2_2_0}.

\section{Theoretical Framework}\label{SEC:Framework}

\subsection{Radiative efficiency of the system}
This section delves into the continuum spectrum emitted by an accretion disc surrounding a black hole, with a specific emphasis on the \emph{Novikov-Thorne model}\cite{Novikov:1973kta,1973blho.conf..343N,1974ApJ...191..499P}. In this model, the accretion disc is assumed to be geometrically narrow, and particles are assumed to follow quasi-circular trajectories that are located close to the equatorial plane. This causes particles to navigate spiral-like patterns before eventually succumbing to gravitational attraction and falling into the black hole. Viscous forces provide small radial motions, which cause particles to navigate towards the black hole.

The examination of circular orbits within the equatorial plane is notably significant. In the Novikov-Thorne model, the accretion disc orbits in a plane that is orthogonal to the spin of the black hole. The gas particles traverse almost geodesic circular trajectories within the equatorial plane. The inner boundary of the disc corresponds to the radius of the \emph{innermost stable circular orbit} (ISCO). 

In light of the fact that it is generally accepted that the gravitational force has a more substantial impact on the motion of gas than the pressure of gas, it is proposed that the particles that are contained within the disc adhere to geodesic orbits that are circular. The gas experiences a decrease in both its energy and its angular momentum as it moves downward into the gravitational well of the central compact object through
the process of gravitational attraction.
Consequently, a fraction of this wasted energy is transformed into electromagnetic radiation.

The Novikov-Thorne accretion disc is defined by its properties of having a small thickness in terms of its geometry i.e \emph{geometrically thin}, being \emph{highly opaque to light}, and lacking any trapped thermal energy. The gas in the disc reaches a state of local thermal equilibrium, causing each location on the disc to display a blackbody spectrum. The combined set of emissions from the disc exhibits a spectrum resembling that of a blackbody at many temperatures. More detailed description can be found in \cite{1973blho.conf..343N, 1974ApJ...191..499P}. In the case of stellar-mass black holes, the emissions tend to reach their highest intensity in the soft X-ray range, whereas for supermassive black holes, the peak intensity occurs in the ultraviolet range.

\subsubsection{Orbits in the equatorial plane}

\begin{figure}[htbp!]
	\centering
	\subfloat[\emph{The radius of Innermost Stable Circular Orbits ($r_{\rm ISCO}$) has been illustrated with variation of spin parameter of the black hole for a set of different values of the dimensionless characteristic parameter/ MOG parameter $\beta$. Solid line designate the radius associated with the Kerr black hole. Whereas, the dashed lines represents radius associated with the regular Kerr-MOG black holes. } ]{{\includegraphics[width=7.5cm]{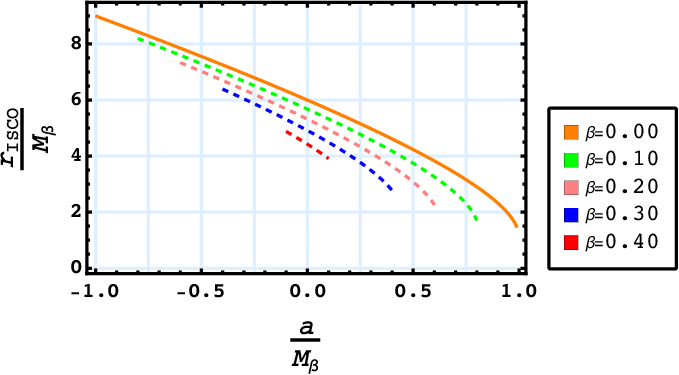}}\label{}}
	\qquad
	\subfloat[\emph{The radius of Innermost Stable Circular Orbits ($r_{\rm ISCO}$) has been shown with change of the dimensionless characteristic parameter $\beta$ of the black hole for a set of various values of the black hole spin parameter $a$. Solid line denotes the radius connected to the non-rotating  black holes. On the other hand, the dashed lines show radius related to rotating black holes.}]{{\includegraphics[width=7.8cm]{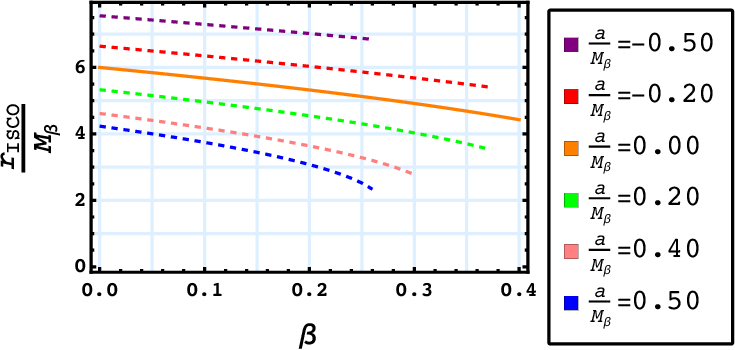}}\label{}}
	\qquad
	\subfloat[\emph{The change in the radius of the  Innermost Stable Circular Orbit ($r_{\rm ISCO}$) has been demonstrated in the parameter-space related to the spin parameter $a$ and the dimensionless characteristic parameter $\beta$. }]{{\includegraphics[width=7.6cm]{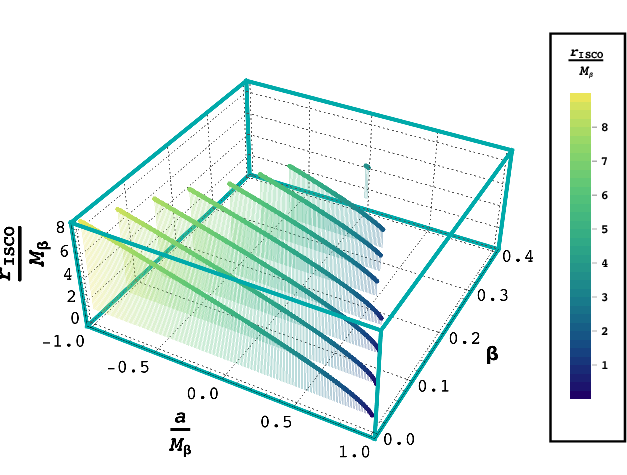}}\label{}}
	\qquad
	\subfloat[\emph{The density-plot of the radius of the  Innermost Stable Circular Orbit ($r_{\rm ISCO}$) has been demonstrated in the parameter-space plane comprising the spin parameter $a$ and the dimensionless characteristic parameter $\beta$. }]{{\includegraphics[width=7cm]{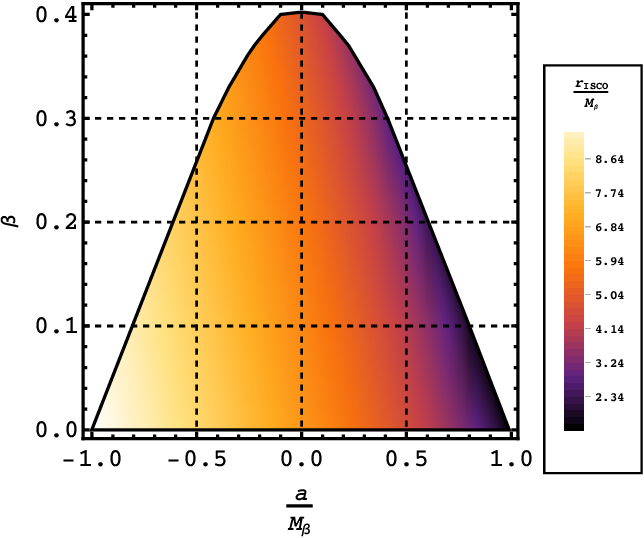}}\label{}}
	\caption{\textit{ Variation of the radius of  Innermost Stable Circular Orbit $\left(r_{\rm ISCO}\right)$ has been depicted with the variation of the dimensionless characteristic parameter $\beta$ and spin parameter $a$ of the black hole. Both prograde and retrograde orbits have been considered in the analysis.}}
	\label{Fig:RISCO2}
\end{figure}

Inner edge of the accretion disc plays a crucial role in the properties of the thermal spectrum originating from the disc. In order to determine the ISCO radius, one need independent measurement of the mass and distance of the compact object along with the inclination angle of the disc. It should be noted that the ISCO radius depends on the background geometry. The effective potential associated with the orbit of the particle around the black hole helps to determine the ISCO radius. In an axisymmetric spacetime, the normalisation condition of particle velocity (given by $u^{\alpha}$), i.e., $u_{\alpha}u^{\alpha}=-1$ leads to
\begin{flalign}
	g_{rr}u_{r}^{2}+g_{\theta\theta}u_{\theta}^{2}=V_{\rm eff}(r,\theta)
\end{flalign}
where $V_{\rm eff}$ is the effective potential of the particle that can be detremined by the following equation\cite{Bardeen1972RotatingBH,Bambi,Novikov:1973kta} 
\begin{flalign}
V_{\rm eff}=\dfrac{\varepsilon^{2}g_{\phi\phi}+\ell^{2}g_{tt}+2\varepsilon\ell g_{t\phi}}{g_{t\phi}^{2}-g_{tt}g_{\phi\phi}}-1\label{Eq:Veff}
\end{flalign}
Here $\varepsilon$ and $\ell$ denotes the specific energy and specific angular momentum of the massive particle, orbitting the compact object. These constants of motion can be expressed in terms of the metric components. The relations are given as
\begin{subequations}
\begin{flalign}
\varepsilon&=-\dfrac{\left(g_{tt}+\Omega_{\rm p} g_{t\phi}\right)}{\sqrt{-g_{tt}-2\Omega_{\rm p} g_{t\phi}-\Omega_{\rm p}^{2}g_{\phi\phi}}}\\
\ell&=\dfrac{\left(g_{t\phi}+\Omega_{\rm p} g_{\phi\phi}\right)}{\sqrt{-g_{tt}-2\Omega_{\rm p} g_{t\phi}-\Omega_{\rm p}^{2}g_{\phi\phi}}}
\end{flalign}
\end{subequations}
where $\Omega_{\rm p}=\dfrac{\dd\phi}{\dd t}$ is the angular velocity of the particle. This is provided explicitly as\cite{Bardeen1972RotatingBH,Bambi,Banerjee:2019sae,Narzilloev:2024fsw}
\begin{flalign}
\Omega_{\rm p\pm}=\dfrac{\dd \phi}{\dd t}=\dfrac{-\partial_{r}g_{t\phi}\pm \sqrt{\left(\partial_{r}g_{t\phi}\right)^{2}-\left(\partial_{r}g_{\phi\phi} \right)\left( \partial_{r}g_{tt}\right)}}{\partial_{r}g_{\phi\phi}}
\end{flalign}
Here, $\Omega_{\rm p+}$ and $\Omega_{\rm p-}$  is associated with the co-rotating i.e angular momentum parallel to the spin of the central gravitating object and counter-rotating orbits i.e angular momentum anti-parallel to the spin of the central gravitating object, respectively.
The radius of the innermost stable circular orbit is connected with to the inflection point of this effective
potential given in \ref{Eq:Veff}. The radius can be determined by solving the following set of equations
\begin{subequations}
\begin{flalign}
V_{\rm eff}(r_{\rm ISCO})&=0\\
\pdv{V_{\rm eff}(r)}{r}\bigg\vert_{r_{ISCO}}&=0\\
\pdv[2]{V_{\rm eff}(r)}{r}\bigg\vert_{r_{\rm ISCO}}&=0
\end{flalign}
\end{subequations}
Notably, by measuring the ISCO from the continuum spectrum, these criteria allow one to constrain the parameters that describe the spacetime surrounding the central compact object. These conditions involve the components of the spacetime metric. For example, the black hole spin can be estimated if the spacetime metric follows the Kerr solution\cite{Zhang:1997dy}. This methodology is often referred to as the \emph{Continuum Fitting Method (CFM)}, extensively applied in literature for spin estimation of stellar-mass black holes\cite{McClintock:2013vwa}. In this context, it is needless to say that we need to know the parametric dependence of ISCO radius. In \ref{Fig:RISCO2}, the parametric dependence of the ISCO radius has been depicted.

\subsubsection{Radiative Efficiency}
\begin{figure}[htbp!]
	\centering
	\subfloat[\emph{Variation of the radiative efficiency $\beta_{\rm NT}$ has been illustrated with the change of the spin parameter $a$ of the black hole for a set of different values of dimensionless characteristic parameter $\beta$. Solid line indicate the radiative efficiency for Kerr black hole, whereas the dashed lines represents the regular black hole.} ]{{\includegraphics[width=7.5cm]{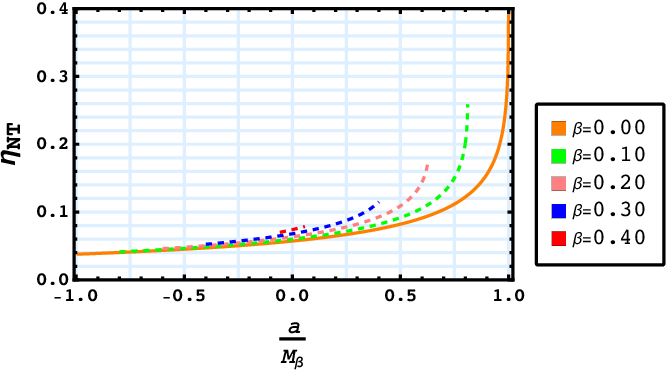}}\label{Fig:TSS_1}}
	\qquad
	\subfloat[\emph{Close view of the variation of the radiative efficiency $\eta_{\rm NT}$ has been illustrated with the change of the spin parameter $a$ of the black hole for a set of different values of dimensionless characteristic parameter $\beta$.}]{{\includegraphics[width=7.5cm]{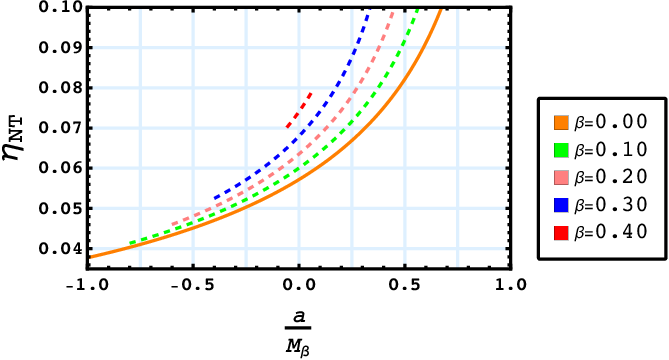}}\label{Fig:TSS_2}}
	\caption{\emph{The change in radiative efficiency $\eta_{\rm NT}$ has been depicted with the variation of the spin parameter $a$ of the black hole for a set of different values of dimensionless characteristic parameter $\beta$. Maximum efficiency can be obtained for the prograde motion of the particle. Maximum value of radiative efficiency can be extracted for Kerr black hole than the same for regular black holes.}}
	\label{Fig:TSS_Variation}
\end{figure}

\begin{figure}
	\centering
	\includegraphics*[scale=1]{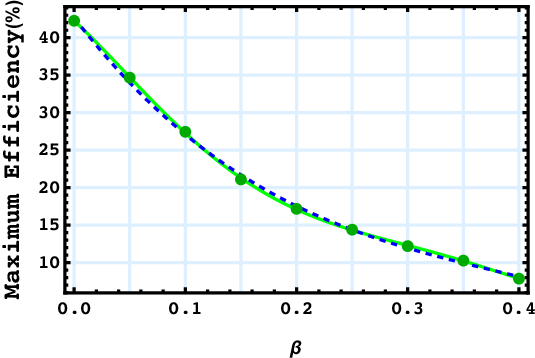}
	\caption{\emph{The maximum efficiencies of  regular Kerr-MOG black hole for various values of dimensionless  parameter $\beta$ have been displayed with the green dots. The green solid  line and dashed blue line  joining the dots has been produced by fitting polynomial of order 6 and 3, respectively.  } }\label{FIG:FIT_MRE}
\end{figure}

Novikov-Thorne accretion disc is usually charcaterized by the binding energy of a particle, orbitting around the black hole. The details of the binding energy is encoded in the \emph{radiative efficiency} $\eta_{\rm NT}$. This is given by
\begin{flalign}
\eta_{\rm NT}=1-\varepsilon_{\rm ISCO}\label{Eq:Rad}
\end{flalign}	
where $\varepsilon_{\rm ISCO}$ represents the specific energy of the particle orbiting at ISCO radius. Therefore, the value of $\eta_{\rm NT}$ depends on the properties of the spacetime metric. For our scenario, it has a  dependency on both the characteristic parameter $\beta$ and spin parameter $a$. In \ref{Fig:TSS_Variation}, we have shown the dependence of the radiative efficiency on the spin parameter of the black hole for a set of fixed values of characteristic parameter $\beta$. The effect of MOG parameter on the maximum radiative efficiency of the black hole has been provided in tabular form in \ref{Tab:RE_MOG}. One  can observe that the maximum efficiency is inversely proportional to the MOG parameter as the increasing value of the parameter narrows down the allowed spin parameter of the black hole. The data for maximum radiative efficiency with change in MOG parameter can be further fitted with polynomial curves as shown in \ref{FIG:FIT_MRE}.

\begin{table}[h]
	\setlength{\arrayrulewidth}{0.2mm}
	\centering
	\begin{tabular}{|c|ccc|}
		\hline
		\rowcolor{Blue!20}Sl.No. & \multicolumn{1}{|c|}{\bf  characteristic parameter $\beta$ }& \multicolumn{1}{|c|}{\bf  Maximum spin value ($a_{\rm max}/M_{\beta}$)} & \multicolumn{1}{|c|}{\bf Maximum Efficiency(\%)} \\ \hline
		i.                               &        \multicolumn{1}{|c|}{0.00}                         &             \multicolumn{1}{|c|}{0.9999999999999}  &     \multicolumn{1}{|c|}{42.2649}              \\ \hline
			\rowcolor{green!20}ii.                          &       \multicolumn{1}{|c|}{0.05}                         &    \multicolumn{1}{|c|}{0.9036537696640} &     \multicolumn{1}{|c|}{34.6742}                       \\ \hline
		iii.                           &      \multicolumn{1}{|c|}{0.10}        &\multicolumn{1}{|c|}{0.8112002347086}                  &        \multicolumn{1}{|c|}{27.4579}                          \\ \hline
	\rowcolor{green!20}	iv.                           &       \multicolumn{1}{|c|}{0.15}                         &    \multicolumn{1}{|c|}{0.7194209939163} &     \multicolumn{1}{|c|}{21.0956}                        \\ \hline
		v.                           &       \multicolumn{1}{|c|}{0.20}                         &    \multicolumn{1}{|c|}{0.6258308679385} &     \multicolumn{1}{|c|}{17.1827}                        \\ \hline
	\rowcolor{green!20}	vi.                           &       \multicolumn{1}{|c|}{0.25}                         &   \multicolumn{1}{|c|}{0.5275250794070} &     \multicolumn{1}{|c|}{14.4005}                      \\ \hline
		vii. &    \multicolumn{1}{|c|}{0.30}                            &       \multicolumn{1}{|c|}{0.4197147218306}    & \multicolumn{1}{|c|}{12.2072}                      \\ \hline
	\rowcolor{green!20}	viii. &    \multicolumn{1}{|c|}{0.35}                            &       \multicolumn{1}{|c|}{0.2908271748714}    & \multicolumn{1}{|c|}{10.2777}                      \\ \hline
		ix.&\multicolumn{1}{|c|}{0.40}                               &  \multicolumn{1}{|c|}{0.0575648519258} &       \multicolumn{1}{|c|}{07.8726}                        \\ \hline
	\end{tabular}
\caption{\emph{Maximum efficiencies in percentage have been displayed for various values of dimensionless characteristic parameter $\beta$. For these values of the parameter $\beta$, the approximate maximum allowed spin values of black hole have been provided accordingly. } }\label{Tab:RE_MOG}
\end{table}

\subsection{Relativistic Jets}
There are two types of jets associated with microquasars:
\begin{enumerate}[i.]
	\item Steady and non-relativistic jets, common in the hard state, endure across a broad spectrum of accretion luminosities.
	\item Transient or ballistic jets emerge when the source luminosity nears the Eddington limit during the transition from the hard to soft state. This sort of jet typically demonstrates relativistic characteristics and is believed to originate near the event horizon.
\end{enumerate}
In this particular setting, we concentrate on the second category of jets in order to gain an understanding of the black hole spin and characteristic parameter associated with regular black hole. A coherent model that can explain observations has not yet been found, despite the fact that several attempts have been made to shed light on the mechanism that is responsible for the creation of relativistic transient jets. The process that Blandford and Znajeck proposed for the extraction of energy from a black hole is the one that we use in our investigation. This method, which can be applied to any axisymmetric spacetime metric, involves the generation of relativistic jets that are powered by the rotational energy of the black hole. These jets are powered by the magnetic field of the accretion disc that is carrying the current.

\section{Observational constraints}\label{SEC:OBS}
As given in \ref{Eq:Rad}, the radiative efficiency  is influenced by the spacetime metric, making its measurement essential for estimating or constraining the black hole characteristics within the relevant gravitational theory. We shall thereafter analyse specific celestial objects within the framework of the normal Kerr-MOG spacetime: GRS1124-683, H1743-322, GRS1915+105, GROJ1655-40, A0620-00, and XTEJ1550-564. \ref{Table1} presents the documented measurements of various features of these systems as reported in the literature. The estimations of black hole spin a and the corresponding calculation of the Novikov-Thorne radiative efficiency $\eta_{NT}$ are calculated under the assumption of the Kerr metric.

\begin{table}[h!]

	\begin{center}
		\begin{tabular}{|c|c|c|c|c|c|c|}
			
			\hline
			$\rm BH~ Source$ & $ M (M_\odot)$ & $ D (kpc)$ & $ i^\circ$ & $ a$ & $\eta$ &$ (S_{\nu,0})_{ max,~ 5GHz} (Jy)$  \\
			\hline 
			$\rm A0620-00$ & $\rm 6.61\pm 0.25$ & $\rm 1.06\pm 0.12$ & $\rm 51.0\pm 0.9$ & $\rm 0.12\pm 0.19$ & $\rm 0.061^{+0.009}_{-0.007}$\cite{Gou2010THESO} & $\rm 0.203$ \\ \hline
			$\rm H1743-322$ & $\rm 8.0$ & $\rm 8.5\pm 0.8$ & $\rm 75.0 \pm 3.0$ & $\rm 0.2\pm 0.3$ & $\rm 0.065^{+0.017}_{-0.011}$\cite{Steiner2011THEDI} & $\rm 0.0346$    \\ \hline
			$\rm XTE J1550-564$ & $\rm 9.10\pm 0.61$ & $\rm 4.38\pm 0.5$ & $\rm 74.7\pm 3.8$ & $0.34\pm 0.24$ & $0.072^{+0.017}_{-0.011}$\cite{2011MNRAS.416..941S} & $\rm 0.265$  \\ \hline
			$\rm GRS 1124-683$ & $\rm 11.0^{+2.1}_{-1.4}$ & $\rm 4.95^{+0.69}_{-0.65}$ & $\rm 43.2^{+2.1}_{-2.7}$ & $\rm 0.63^{+0.16}_{-0.19}$ & $\rm 0.095^{+0.025}_{-0.017}$\cite{Chen2015THESO} & $\rm 0.45$  \\ \hline
			$\rm GRO J1655-40$ & $\rm 6.30\pm 0.27$ & $\rm 3.2\pm 0.5$ & $\rm 70.2\pm 1.9$ & $\rm 0.7\pm 0.1$ & $\rm 0.104^{+0.018}_{-0.013}$\cite{Shafee:2005ef} & $\rm 2.42$ \\ \hline
			$\rm GRS 1915+105$ & $\rm 12.4^{+1.7}_{-1.9}$ & $\rm 8.6^{+2.0}_{-1.6}$ & $\rm 60.0\pm 5.0$ & $\rm  a_*>0.98$ & $  \eta>0.234$\cite{McClintock:2006xd} & $\rm 0.912$  \\ \hline
		\end{tabular}
	\end{center}
		\caption{Parameters of the transient black hole binaries}
	\label{Table1}
\end{table}

Spin measurements utilising the continuum fitting approach are acquired when the source is in a soft state, with accretion luminosity ranging from around 5\% to 30\% of the Eddington limit. Under these circumstances, the disc is expected to be accurately characterised by the Novikov-Thorne model, wherein the inner boundary of the disc is situated at the ISCO radius and outflows and jets are considered negligible. 

A bipolar radio jet is conceptualised as a symmetrical dual of plasmoids. These plasmoids emit radiation isotropically and exhibit a narrow optical structure, extending outward from the core source at a relativistic bulk velocity $\widetilde{\beta}$. The ratio of observed to emitted flux density for each jet can be provided as\cite{Steiner_2013,1994Natur.371...46M}
\begin{flalign}
\dfrac{S_{\nu}}{S_{\nu,0}}=\delta^{3-\alpha}
\end{flalign}
Here $\delta$ denotes the Doppler factor and $\alpha$ represents the radio spectral index. The Doppler factor for the more luminous jet, moving towards the observer, can be succinctly articulated in relation to $\tilde{\beta}$, the Lorentz factor $\Gamma$, and the inclination angle $i$ of the jet
\begin{flalign}
\delta=\left[\Gamma \left(1-\widetilde{\beta}\cos i  \right) \right]^{-1}
\end{flalign}
For the principal emission source, specifically the incoming jet, the measured intensity exceeds the emitted intensity at lower angles, whereas the opposite holds true at higher angles. For microquasars with slightly relativistic jets, the Doppler boost diminishes to below one within the intermediate range of inclinations, namely between 35 and 55 degrees. We assume that the total power in the transient jet is proportional to the peak radio flux density at 5 GHz. In natural units, the the jet power can
be given by\cite{Narayan:2011eb,Steiner2012JETPA}
\begin{flalign}
P_{\rm jet}=\left(\dfrac{\nu}{5~\rm GHz} \right)\left(\dfrac{S_{\nu,0}^{\rm tot}}{\rm J ~y} \right)\left(\dfrac{D}{\rm kpc} \right)^{2} \left(\dfrac{M}{M_{\odot}} \right)^{-1}
\end{flalign}

where, $\nu S_{\nu,0}^{\rm tot}$ is the beaming corrected maximum flux after taking into account the approching and receding jets\cite{Steiner2012JETPA,Mirabel:1999fy}. The Lorentz factor $\Gamma$ connected to the jet is projected to lie within the range $2\lesssim \Gamma \lesssim 5$. 
\ref{Table2} provides the Doppler-corrected jet powers for each source that correspond to Lorentz factors $\Gamma=2$ and $\Gamma=5$\cite{Pei:2016kka,Middleton:2014cha}. The outputs reported in \ref{Table2} can be compared with the theoretical predictions, dependant on the space- time metric. 

\begin{table}[h!]
	\begin{center}
		\begin{tabular}{|c|c|c|}
			
			\hline
			$\rm BH~ Source$ & ${\rm  P_{jet}}\rvert_{\Gamma=2} $ & ${\rm P_{jet}}\rvert_{\Gamma=5}$  \\
			\hline 
			$\rm A0620-00$ & $\rm 0.13$ & $\rm 1.6$  \\ \hline
			$\rm H1743-322$ & $\rm 7.0$ & $\rm 140$  \\ \hline
			$\rm XTE J1550-564$ & $\rm 11$ & $\rm 180$  \\ \hline
			$\rm GRS 1124-683$ & $\rm 3.9$ & $\rm 390$  \\ \hline
			$\rm GRO J1655-40$ & $\rm 70$ & $\rm 1600$  \\ \hline
			$\rm GRS 1915+105$ & $\rm 42$ & $\rm 660$  \\ \hline

		\end{tabular}\label{Tab:DoP25}
		
	\end{center}
		\caption{Proxy jet power values in units of $\rm kpc^2~GHz ~Jy ~M_\odot^{-1}$}
	\label{Table2}
\end{table}

From \ref{Eq:JetPow}(see \ref{App_2_2}), the jet’s power can be stated as
\begin{flalign}
	&\log P= \log K +\log \left[1-F(r_{\rm H}) \right] + 2\log \Omega_{\rm H}\\
	\implies& \log P=\log K+ 2\log \sqrt{1-F_{H}}\Omega_{H}
\end{flalign}
where $K=k\Phi_{\rm tot}^{2}$. The calculation of K entails fitting the observed jet power and $\Omega_{H}$\cite{Narayan:2011eb,Middleton:2014cha}.
The spacetime metric affects jet power calculations via the square of the angular velocity at the event horizon $\Omega_{H}^{2}$. In the study referenced as \cite{Middleton:2014cha}, the authors determined the optimal values for the parameter $K$. The given values are $\log K = 2.94 \pm 0.22$ for the Lorentz factor $\Gamma=2$, and $\log K = 4.19 \pm 0.22$ for $\Gamma=5$, at a 90\% confidence level. $K$ may not be universally constant for all sources. It is hypothesised that the strength of the magnetic field is contingent upon the mass accretion rate\cite{Narayan:2011eb,10.1111/j.1745-3933.2011.01147.x}.

As transient jets emerge during the transition from the hard to soft state, the Eddington-scaled mass accretion rate remains consistent across all sources. Furthermore, the masses of all objects are comparable, approximately $10~M_{\odot}$. Consequently, this quantity may be regarded as constant for the six black hole possibilities in our enumeration.
Assuming $K$ is independent of the spacetime geometry, we utilise the established values of $K$ to impose constraints on the spin parameter and the Kerr-MOG regularity parameter $\beta$, based on the observed jet power of the sources listed in \ref{Table2}.

\subsection{observational estimations}
\subsubsection{A0620-00}
The X-ray binary A0620-00 consists of a K-type main sequence star and a black hole with a mass of 6.6 $M_{\odot}$\cite{2010ApJ...710.1127C}. It is the closest identified X-ray binary to the solar system\cite{Foellmi:2008gr}, possessing an orbital period of 7.75 hours\cite{1986ApJ...308..110M,2010ApJ...718L.122G}. The distance and angle of the source are presented in \ref{Table1}\cite{2010ApJ...710.1127C}. The source's dimensionless spin parameter ($\tilde{a}$) has been ascertained using the Continuum-Fitting method, yielding $-0.59<\tilde{a}<0.49$, with the optimal value being $\tilde{a}=0.12\pm0.19$\cite{2010ApJ...718L.122G}, allowing for the calculation of its radiative efficiency $\eta_{NT}$.

\begin{figure}[htbp!]
	\centering
	\subfloat[\emph{} ]{{\includegraphics[width=6cm]{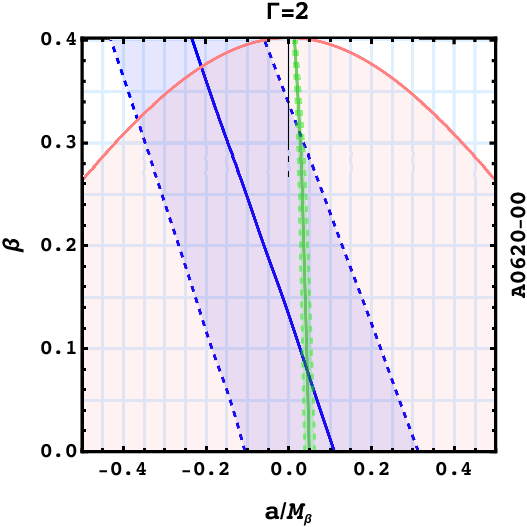}}\label{Fig:PE_1a}}
	\qquad
	\subfloat[\emph{}]{{\includegraphics[width=6cm]{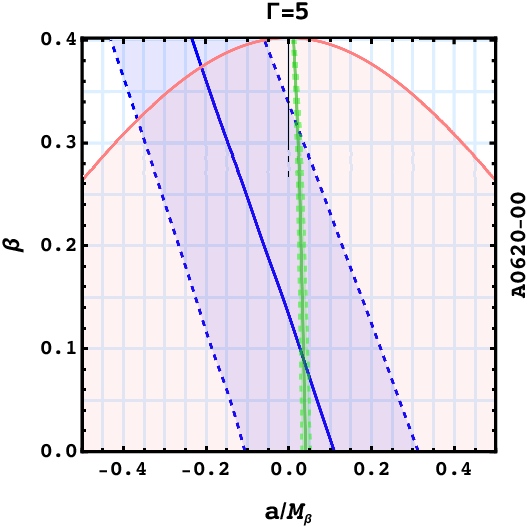}}\label{Fig:PE_1b}}
	\caption{\emph{Constraints on the spin parameter $a$ and the MOG parameter $\beta$ for the black hole candidate A0620--00 has been shown. The blue shaded region corresponds to parameter values that yield a radiative efficiency consistent with the observed value $\eta_{NT} = 0.061^{+0.009}_{-0.007}$. The green shaded area represents combinations of $(a, \xi)$ that reproduce the measured jet power within an observational uncertainty of $0.3\, \text{dex}$. The intersection of these two regions identifies the subset of the $(a, \xi)$ parameter space that is compatible with both observables in the context of our concerned theory. The red shaded zone denotes the theoretically permitted region for a regular black hole solution.
	}}
	\label{Fig:Parameter_Estimation_1}
\end{figure}

The blue shaded area in \ref{Fig:Parameter_Estimation_1} delineates the permissible values of $\beta$ and $a/M_{\beta}$ that can account for the radiative efficiency of this source, considering the error margins. The blue solid line represents the contour in the $\beta-a/M_{\beta}$ plane where the theoretical radiative efficiency, as described by \ref{Eq:Rad}, aligns with the central value of the observed (\ref{Table1}). The blue dashed lines are likewise correlated with the error bars in the observed $\eta_{NT}$.

Radio investigations of the object indicate the existence of powerful radio jets\cite{2010ApJ...718L.122G,Kuulkers:1999kn}, with a $5~\rm  GHz$ radio flux density of $0.203\rm ~ Jy$\cite{Narayan:2011eb} (\ref{Table2}). The previously mentioned radio flux density is transformed into radiated radio luminosity using Doppler deboosting with Lorentz factors $\Gamma=2$ and $\Gamma=5$, with the proposed values detailed in \ref{Table2}. Theoretical jet power $P_{BZ}$ (as defined by \ref{Eq:JetPow}) is subsequently compared to identify the permissible values of the parameters $\beta$ and $a$ based on jet-related observations. An error of $0.3~\rm dex$ is taken in the measured jet power $P_{\rm jet}$\cite{Narayan:2011eb,Middleton:2014cha}. The green shaded area in \ref{Fig:Parameter_Estimation_1} illustrates the permissible values of $\beta$ and $a$ that can account for the measured jet power within the error margins. The solid green line illustrates the contour in the $\beta$-$a$  plane that reproduces the central value of $P_{\rm jet}$, while the dashed green lines denote the values of $\beta$ and $a$ that account for the jet power within an error margin of $0.3~\rm dex$ from the central value.

The outcomes for $\Gamma=2$ and $\Gamma=5$ are illustrated in \ref{Fig:PE_1a} and \ref{Fig:PE_1b}, respectively. The red coloured area is the parameter space in the $\beta$-$a$ plane corresponding to true positive event horizons that result in black hole solutions within MOG gravity. In the ensuing discussion, the definitions of the red, blue, and green shaded regions stay consistent for the other X-ray binaries. For relatively high values of parameter $\beta$, the counter-rotating orbits are more favoured to generate specific radiative efficiency. However, from the plots it is clear that theoretically allowed extreme value of $\beta$ can not produce the required radiative efficiency for this system. Conversely, the jet power can be replicated by nearly the complete spectrum of the parameter $\beta$. The overlapping of the blue and green shaded areas denotes the permissible values of $\beta$ and $a$ that can account for both observations concerning $P_{\rm jet}$ and $\eta_{\rm NT}$. \ref{Fig:Parameter_Estimation_1} indicates that the range $0\lesssim \beta\lesssim 0.395$ can account for both previously mentioned observations. Furthermore, the permissible ranges of spin from both measurements demonstrate an intersection within the framework of general relativity ($\beta= 0$).

\subsubsection{H1743-322}

This galactic microquasar is situated at a distance of $8.5\pm0.8~\rm kpc$ and possesses an inclination of $75^{\circ} \pm 3^{\circ}$\cite{Steiner2011THEDI}. The mass of this object has not been dynamically measured; nonetheless, it is anticipated to vary between $8~M_{\odot}$ and $13~M_{\odot}$\cite{Pei:2016kka,Petri:2008jc}. The companion star is a late-type main sequence star situated in the galactic bulge\cite{Chaty:2015pda}, with an orbital period of $10$ hours for the binary system\cite{2010MNRAS.401.1255J}. The object's spin, as determined using the Continuum-Fitting approach, is $0.2 \pm 0.3$ at $68\%$ confidence and $-0.3 < a < 0.7$ with $90\%$ confidence\cite{Steiner2011THEDI}. The associated radiative efficiency is presented in \ref{Table1}. 

\begin{figure}[htbp!]
	\centering
	\subfloat[\emph{} ]{{\includegraphics[width=6cm]{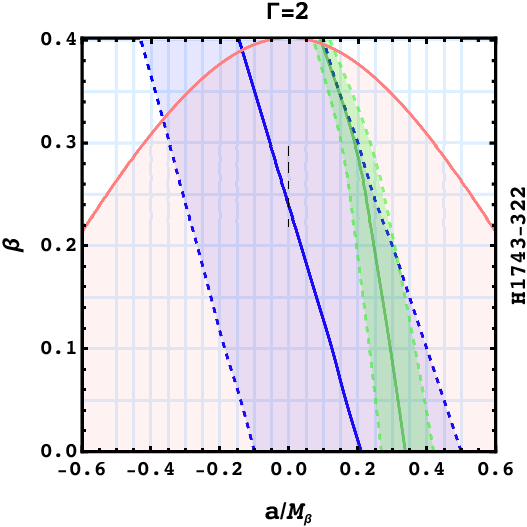}}\label{Fig:PE_2a}}
	\qquad
	\subfloat[\emph{}]{{\includegraphics[width=6cm]{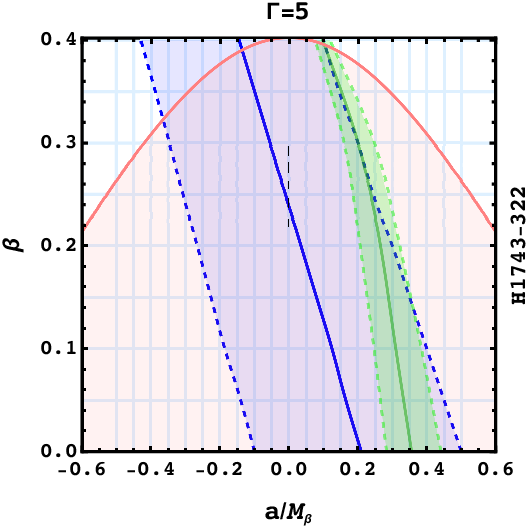}}\label{Fig:PE_2b}}
	\caption{\emph{Jet power and radiative efficiency constraints for the black hole system H1743--322 are illustrated. In each panel, the green shaded region indicates the combinations of the MOG parameter $\beta$ and spin $a$ that successfully reproduce the observed jet power, accounting for uncertainties, for (a) Lorentz factor $\Gamma = 2$ and (b) $\Gamma = 5$. The blue shaded region represents the parameter space consistent with the observed radiative efficiency $\eta_{NT}$, as summarized in \ref{Table1}. A comprehensive discussion of the astrophysical implications of these overlapping and distinct regions is presented in the main text.
	}}
	\label{Fig:Parameter_Estimation_2}
\end{figure}

The blue shaded area in \ref{Fig:Parameter_Estimation_2} corresponds to the permissible values of $\beta$ and $a$ that can account for the reported radiative efficiency within the error margins. The blue lines represent the contours in the $\beta$-$a$ plane when the observed $\eta_{NT}$ is replicated by the theoretical radiative efficiency as specified in \ref{Eq:Rad}, with the solid blue line indicating the central value and the dotted blue lines illustrating the errors surrounding the central value in \ref{Table1}. From the plot, it is evident that the whole range of parameter $\beta$ can explain the radiative efficiency.

The cosmic object displays strong ballistic jets\cite{Steiner2011THEDI}, with the emitted jet power for $\Gamma=2$ and $\Gamma=5$ detailed in \ref{Table1}. These are linked to an error of $0.3~\rm dex$ relative to the central value\cite{Narayan:2011eb,Middleton:2014cha}. \ref{Fig:Parameter_Estimation_2} illustrates the permissible values of $\beta$ and $a$ that account for the emitted jet power, represented by the green shaded region, which includes both positive and negative errors. The definition of the blue solid and dashed lines remains unchanged from the prior scenario. The emitted jet power for $\Gamma=2$ and $\Gamma=5$ is illustrated in \ref{Fig:PE_2a} and \ref{Fig:PE_2b}, respectively. We additionally observe that nearly $\beta\lesssim 0.39$ can characterise the emitted jet power, and the  measured radiative efficiency. The intersection area between the blue and green shaded regions signifies the values of $\beta$ and $a$ that characterise both observations.

\subsubsection{XTE J1550-564}

\begin{figure}[htbp!]
	\centering
	\subfloat[\emph{} ]{{\includegraphics[width=6cm]{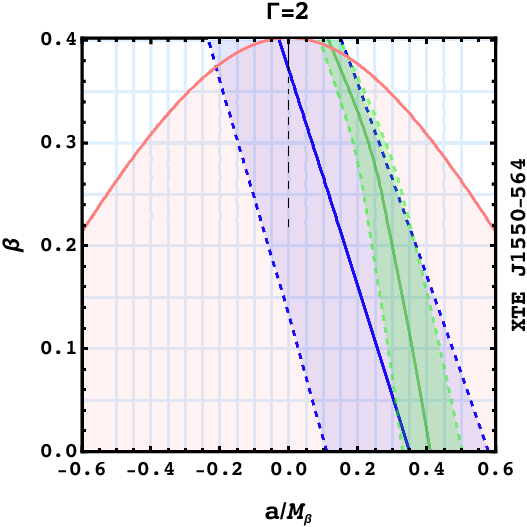}}\label{Fig:PE_3a}}
	\qquad
	\subfloat[\emph{}]{{\includegraphics[width=6cm]{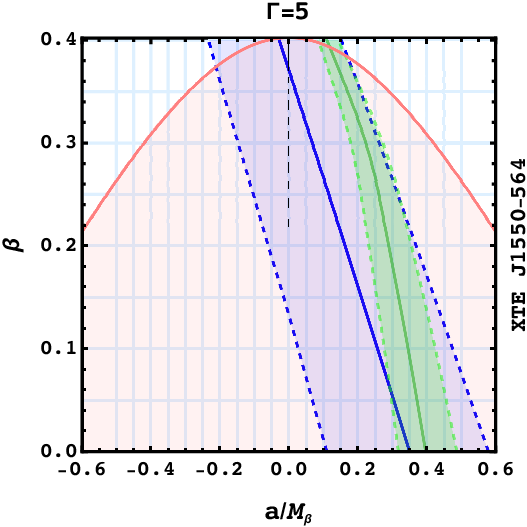}}\label{Fig:PE_3b}}
	\caption{\emph{Parameter constraints for the black hole binary XTE J1550$-$564 are presented. The green shaded regions in panels (a) and (b) depict the combinations of the MOG parameter $\beta$ and spin $a$ that yield theoretical jet power in agreement with the observed radio luminosity, after accounting for Doppler boosting with Lorentz factors $\Gamma = 2$ and $\Gamma = 5$, respectively. The blue shaded region indicates the portion of the parameter space where the model-predicted radiative efficiency matches the observational estimate reported in \ref{Table1}. The red shaded zone delineates the values of $\beta$ and $a$ that admit a regular event horizon, confirming the presence of a regular black hole solution within the framework. Further physical interpretation and implications of these constraints are elaborated in the main text.
	}}
	\label{Fig:Parameter_Estimation_3}
\end{figure}

XTE J1550-564 is a binary system with a black hole with a mass of $9.1\pm 0.61~M_{\odot}$\cite{2011ApJ...730...75O} and a companion star of late G or early K spectral class\cite{Orosz:2001qd}. The  orbital period of the binary is 1.55 days\cite{Orosz:2001qd}.  The distance and inclination of the source are $4.38^{+0.58}_{-0.41}~\rm  kpc$ and $74.7^{\circ} \pm 3.8^{\circ}$, respectively\cite{2011ApJ...730...75O}.  The black hole's spin has been assessed using both the Continuum Fitting and Fe-line methods. The outcome derived from the Continuum Fitting approach indicates $-0.11 < a < 0.71$ ($90\%$ confidence)\cite{2011MNRAS.416..941S}, with the most probable spin being $a = 0.34$, but the Fe-line method yields a spin estimate of $a = 0.55^{+0.15}_{-0.22}$\cite{2011MNRAS.416..941S}. \ref{Table1} presents the spin associated with the Continuum Fitting technique, and $\eta_{NT}$ is computed based on this outcome\cite{Pei:2016kka}. The object shows a $5\rm  GHz$ radio flux density of $0.265\rm  Jy$.  The emitted jet powers are estimated using Lorentz factors $\Gamma = 2$ and $\Gamma = 5$, as detailed in \ref{Table2}. An inaccuracy of $0.3\rm dex$ is linked to the reported jet powers. The emitted jet powers (corresponding to $\Gamma=2$ and $\Gamma=5$) and their associated errors are compared with the theoretical jet power, with findings illustrated in \ref{Fig:PE_3a} and \ref{Fig:PE_3b}, respectively. The values of $\beta$ and $a$ that account for the emitted jet power within the error margins are represented by the green shaded region.  The blue shaded area, conversely, depicts the permissible values of $\beta$ and $a$ when the theoretical radiative efficiency matches the measured one. As previously noticed, $0.385\lesssim\beta$ cannot account for the observed $\eta_{NT}$ or the measured jet power. 

\subsubsection{GRS 1124-683}
\begin{figure}[htbp!]
	\centering
	\subfloat[\emph{} ]{{\includegraphics[width=6cm]{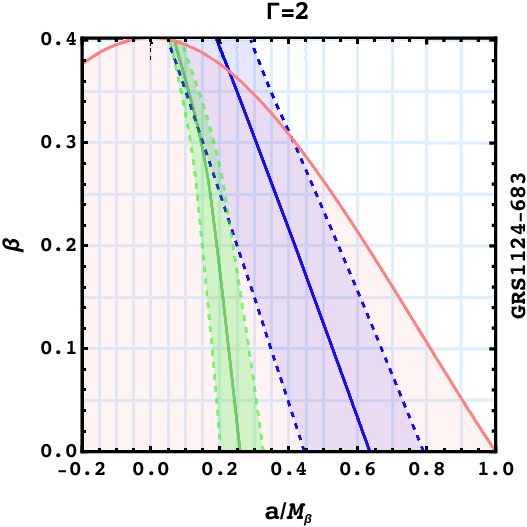}}\label{Fig:PE_4a}}
	\qquad
	\subfloat[\emph{}]{{\includegraphics[width=6cm]{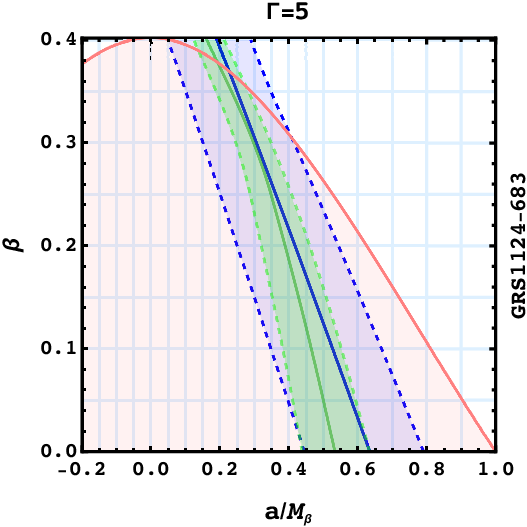}}\label{Fig:PE_4b}}
	\caption{\emph{Parameter space analysis for the black hole candidate GRS 1124$-$683 has been performed. Panel (a) demonstrates that for a jet Lorentz factor $\Gamma = 2$, there exists narrow region in the $(a, \beta)$ parameter space that simultaneously accounts for both the observed jet power and the radiative efficiency. In contrast, panel (b), corresponding to $\Gamma = 5$, reveals an overlap region where both observational constraints are satisfied. This outcome highlights the importance of high-velocity jet ejection in reconciling the system’s radiative and kinetic signatures. A detailed physical interpretation of these findings is provided in the main text.
	}}
	\label{Fig:Parameter_Estimation_4}
\end{figure}

This X-ray binary consists of a black hole with a mass of $11.0^{+2.1}_{-1.4}M_{\odot}$\cite{Wu:2015ioy} and a K-type main sequence star as its companion, exhibiting an orbital period of $10.4~\rm hours$\cite{1992ApJ...399L.145R}. The distance to the source is $D = 4.95^{+0.69}_{-0.65}~\rm kpc$, and the inclination is $i = 43.2^{+2.1^{\circ}}_{-2.7} $\cite{Wu:2015ioy}. The object's spin has been approximated using the Continuum Fitting method, yielding $a = 0.63^{+0.16}_{-0.19}$\cite{Chen2015THESO}. The radiative efficiency has been determined based on this value of the Kerr parameter (\ref{Table1}).
The permissible values of $\beta$ and a derived from the measured $\eta_{NT}$ are delineated by the blue shaded area in \ref{Fig:Parameter_Estimation_4}, indicating that $\beta_{\rm max}$ is approximately 0.4. The emitted jet power for this source at $\Gamma=2$ and $\Gamma=5$ is presented in \ref{Table2}.
An uncertainty of $0.3~\rm dex$ is assumed for the reported power\cite{Narayan:2011eb,Middleton:2014cha}.

In \ref{Fig:Parameter_Estimation_4}, the green shaded area delineates the permissible values of $\beta$ and a that can account for the emitted jet power within the error margins. For $\Gamma=2$, there is a very small overalapping region or parameter space where both the jet power and radiative efficiency can be simultaneously reproduced. However, interestingly for $\Gamma=2$, the entire parameter space that captures the region for successful reproduction of jet power is a subset of the region that reproduces the desired radiative efficiency. For the later scenario, the maximum allowed parameter value will be $\beta\lesssim 0.38$. Furthermore, in contrast to earlier black holes, the permissible spin range that may account for observations at $\beta=0$ exhibits overlap just when $\Gamma=5$ is utilised to calculate the emitted jet power from the observed $5~\rm GHz$ radio-flux density.

\subsubsection{GRO J1655-40}

\begin{figure}[htbp!]
	\centering
	\subfloat[\emph{} ]{{\includegraphics[width=6cm]{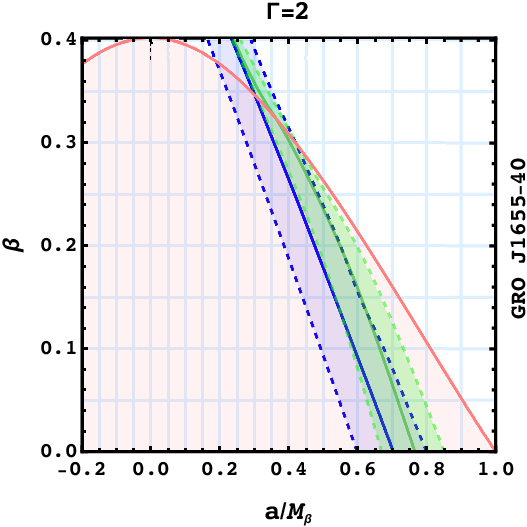}}\label{Fig:PE_5a}}
	\qquad
	\subfloat[\emph{}]{{\includegraphics[width=6cm]{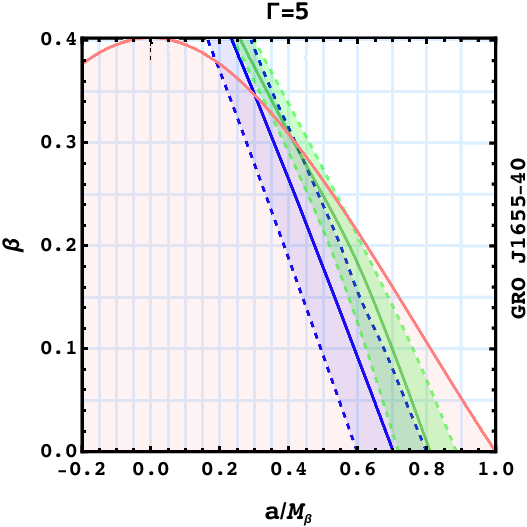}}\label{Fig:PE_5b}}
	\caption{\emph{Parameter constraints for GRO~J1655$-$40 have been depicted. The 'panel(a)' and 'panel (b)' correspond to Lorentz factors $\Gamma = 2$ and $\Gamma = 5$, respectively. In both cases, regular rotating black hole in MOG theory provides a consistent description of the observed radiative efficiency $\eta_{NT}$ and the jet power $P_{\mathrm{jet}}$. A more detailed explanation is provided in the main text.}}
	\label{Fig:Parameter_Estimation_5}
\end{figure}

GRO J1655-40 comprises a black hole with a dynamical mass of $M = \left(6.3 \pm 0.5\right) M_{\odot}$\cite{Greene:2001wd} and an F-type secondary star with a mass of $M_{star} = \left(2.34 \pm 0.12\right) M_{\odot}$, exhibiting an orbital period of $2.62~ \rm days$\cite{Orosz:1996cg}.
The distance to the source is estimated at $D = \left(3.2 \pm 0.5\right)~\rm kpc$\cite{Hjellming:1995tv}, and its orbital inclination is $i = 70.2^{\circ} \pm 1.9^{\circ}$\cite{Greene:2001wd}. There is considerable disagreement around the bias of this source. The Continuum Fitting approach forecasts a spin of approximately $0.65$ to $0.75$\cite{Shafee:2005ef}, but the spin determined using the Fe-line method exceeds $0.9$\cite{2009MNRAS.395.1257R}. The quasi-periodic oscillations detected in the power spectrum of GRO J1655-40 have limited its mass and spin to $M = \left(5.31 \pm 0.07\right) M_{\odot}$ and $a = 0.290 \pm 0.003$, respectively\cite{Motta:2013wga}. In this study, we utilise the spin determined using the Continuum Fitting approach to assess the radiative efficiency. As previously stated, the permissible values of $\beta$ and $a$ that can account for the observed $\eta_{NT}$ within the error margins are highlighted in blue in \ref{Fig:Parameter_Estimation_5}, indicating that $\beta_{\rm max}$ is approximately 0.38.

The $5 ~\rm GHz$ radio-flux density of $2.42~\rm Jy$ has been utilised to assess the emitted jet power, assuming Lorentz factors $\Gamma=2$ and $\Gamma=5$, as detailed in \ref{Table2}. These are linked to an inaccuracy of $0.3~\rm dex$. In \ref{Fig:Parameter_Estimation_5}, the green shaded area denotes the values of $\beta$ and $a$ that can account for the emitted jet power within the permissible deviations. \ref{Fig:PE_5a} and \ref{Fig:PE_5b} correspond to the emitted jet power calculated with $\Gamma=2$ and $\Gamma=5$, respectively.
The reported jet power can be elucidated by the almost complete spectrum of $\beta$ for $\Gamma=2$. Whereas for $\Gamma=5$, the maximum allowed range of $\beta$ is approximately $0.38$ when jet power is considered. For both the scenario this upper bound of $\beta$, when considered the jet power  is lower than that of the upper bound when radiative efficiency is taken into account.

\subsubsection{GRS 1915+105}
\begin{figure}[htbp!]
	\centering
	\subfloat[\emph{} ]{{\includegraphics[width=6cm]{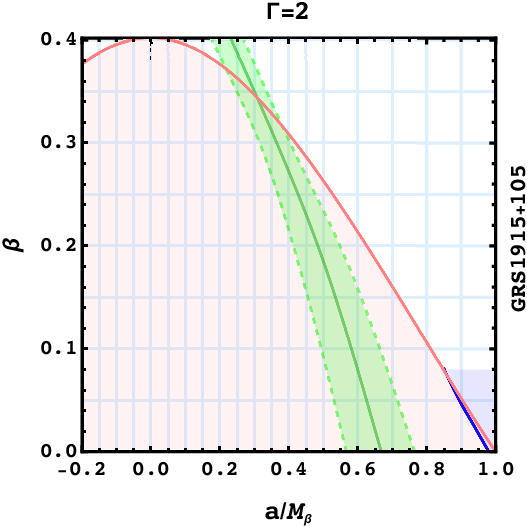}}\label{Fig:PE_6a}}
	\qquad
	\subfloat[\emph{}]{{\includegraphics[width=6cm]{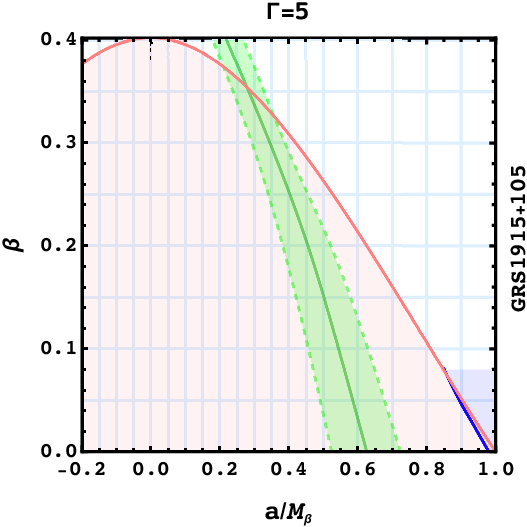}}\label{Fig:PE_6b}}
	\caption{\emph{Analysis of the black hole candidate GRS~1915$+$105 is presented. Panels (a) and (b) indicate that within the framework of a regular rotating spacetime in MOG theory of gravity, it remains challenging to simultaneously reproduce both the observed jet power and the measured radiative efficiency. This tension between theory and observation suggests inherent limitations of the model in describing this particular source. A comprehensive interpretation of these results is discussed in the main text.
	}}
	\label{Fig:Parameter_Estimation_6}
\end{figure}

GRS 1915+105 is a galactic X-ray binary of a black hole and a K-type star, exhibiting an orbital period of 34 days\cite{1994Natur.371...46M,Greiner:2001vb}. The black hole in this X-ray binary possesses a mass of $M = 12.4^{+2.0}_{-1.8} M_{\odot}$\cite{Reid:2014ywa} . The distance to the source is $8.6^{+2.0}_{-1.6} ~\rm kpc$, and the inclination angle is $60^{\circ} \pm 5^{\circ}$\cite{Reid:2014ywa}. The black hole's spin, measured via the Continuum-Fitting method, is more than $0.98$\cite{McClintock:2006xd}, which is utilised to assess radiative efficiency. In \ref{Fig:Parameter_Estimation_6}, the blue shaded area delineated by the blue dashed and solid lines illustrates the permissible values of $\beta$ and $a$ that can account for the observed $\eta_{NT}$. A greater $\beta$ value necessitates a reduced spin to achieve the radiative efficiency. The upper bound of the parameter $\beta$ in this scenario is approximately $0.08$.

The object displays powerful radio jets, with a $5~\rm  GHz$ radio flux density of $0.912~\rm  Jy$\cite{1994Natur.371...46M}. The emitted jet power, calculated from the flux density following Doppler de-boosting with Lorentz factors $\Gamma=2$ and $\Gamma=5$, is presented in \ref{Table2}. The inaccuracy related to the jet power remains $0.3~\rm  dex$.  The green shaded area in \ref{Fig:Parameter_Estimation_6} represents the values of $\beta$ and $a$ that can account for the observed jet power within the error margins.  The solid and dashed green lines retain the same definition as previously stated. There is no common range of parameter space that can explain both the jet power and radiative efficiency.

\section{Conclusions}\label{SEC:CON}

{

The study of rotating black holes surrounded by accretion structures remains a cornerstone of high-energy astrophysics and strong gravity phenomenology. The extraction of rotational energy through magnetohydrodynamic processes, such as those described by the Blandford-Znajeck mechanism, is deeply intertwined with the near-horizon geometry and the angular velocity of field lines. In regular black hole models, where curvature singularities are avoided, the horizon structure is significantly modified, leading to shifts in the innermost stable circular orbits and deviations in the redshift profile and frame-dragging effects. These geometric features determine both the nature of accretion and the jet launching region. Since the efficiency of radiation emitted from the disc and the power transported by jets are closely linked to the location of the ISCO and the angular velocity at the horizon, they provide sensitive diagnostics of the underlying gravitational field. When a compact object possesses both strong magnetic threading and a rapid rotation, the flux of energy extracted becomes a direct function of the deviation from the Kerr limit, especially in theories where additional vector or scalar degrees of freedom are present. This motivates the use of multiwavelength observations as a way to test whether the astrophysical black holes we observe truly conform to the predictions of general relativity or exhibit consistent departures suggestive of modified gravity.

Our results show that the deviation parameter $\beta$, which encapsulates the strength of the extra vector field in MOG theory, leaves clear signatures in the spin-efficiency relation and in the correlation between jet power and black hole angular velocity. Unlike the Kerr solution, where the structure of the ergosphere and ISCO is solely controlled by mass and spin, regular black holes exhibit a more flexible dependence on these quantities, leading to distinct observational consequences. For instance, an enhanced gravitational potential due to the modified field equations can mimic the effects of higher spin, while simultaneously modifying the energetics of jet production. These changes are not just limited to geometry but extend to thermodynamic stability and the topology of causal boundaries. The absence of singularities also allows for the study of quantum effects near the core in a self-consistent manner, which may ultimately tie into a more complete theory of quantum gravity. The study undertaken here shows that constraints arising from radiative and kinetic observables are already powerful enough to rule out large sectors of the allowed parameter space in such models, especially in the high-spin regime where both efficiency and jet power are expected to peak. Therefore, this analysis serves not only as a consistency check for the theoretical models but also as a guide for future observations that aim to test the nature of gravity in the strong-field regime.

In this work, we have systematically analyzed the observed jet powers and radiative efficiencies of several X-ray binary black holes to constrain the allowed ranges of black hole spin and the deviation parameter $\beta$ from the Kerr geometry. Our study employed both continuum fitting and Fe-line spin measurements and compared them with theoretical model predictions for jet production and efficiency. For typical black hole binaries examined, we find that (i) the intersection of constraints from jet power and efficiency imposes an upper bound of $\beta \lesssim 0.4$  for most sources when conservative assumptions for jet Lorentz factors are adopted, (ii) rapidly spinning objects such as GRS 1915+105 demand significantly lower values of $\beta$, indicating a strong sensitivity to deviations in highly relativistic regimes, and (iii) in some cases (notably GRS 1915+105) there is no simultaneous parameter space that can account for both jet power and efficiency, highlighting a limitation of the regular black hole model for extreme spins.

Furthermore, we developed and implemented a modified procedure to generate physically consistent rotating regular black hole solutions, overcoming traditional ambiguities associated with the Newman–Janis algorithm. We also presented concise formulae linking jet power output to black hole angular frequency and magnetic flux in the strong-gravity regime.

Overall, our results provide meaningful observational constraints on possible deviations from Kerr black holes and suggest that current and future accurate measurements of jet power and radiative efficiency, especially in near-extremal black holes, will be crucial in testing the nature of black holes and the underlying gravitational theory in the strong-field regime.

From a broader perspective, the observed interplay between spin-induced frame dragging and the amplification of electromagnetic fields near the black hole horizon suggests a universal mechanism of jet formation across different compact objects. If deviations from the Kerr paradigm exist in nature, their influence will not remain confined to theoretical extensions but will leave observable traces in multi-band electromagnetic spectra and potentially in gravitational wave signatures from black hole binaries. Future missions with improved spectral resolution and timing accuracy, such as the enhanced X-ray Timing and Polarimetry satellite (eXTP) or the Large Observatory For X-ray Timing (LOFT), may enable us to detect subtle shifts in the ISCO radius and test whether the inferred spin-energy correlation remains consistent with general relativity. In addition, very long baseline interferometry (VLBI) imaging of jet structures and black hole shadows may reveal geometrical signatures that deviate from Kerr expectations. The existence of a regular core, for instance, could influence light propagation and photon ring morphology in ways that are yet to be fully quantified. This study, therefore, opens up the possibility of using energetic and geometric observables as a coherent framework to distinguish between singular and nonsingular compact objects, and ultimately to test the foundational assumptions of gravity itself.

}

\section*{Acknowledgement}
SS expresses sincere thanks to Apratim Ganguly for the kind hospitality during the visit to IUCAA, where part of this work was conducted.

\appendix
\section{Appendix}
\subsection{Modified Newman-Janis Algorithm: Derivation of the Rotating Solution from a Static Spacetime}\label{App:NJA}

We present a systematic procedure for deriving a rotating solution in Boyer-Lindquist coordinates (BLCs) starting from a static spherically symmetric spacetime. This approach avoids the ambiguities of the Newman-Janis algorithm (NJA), particularly the arbitrary complexification of coordinates, and instead utilizes symmetry arguments and physically motivated deformations of the static geometry.

Consider a general static spacetime of the form:
\begin{equation}
	\dd s^2 = -G(r)\dd t^2 + \frac{\dd r^2}{F(r)} + H(r)(\dd\theta^2 + \sin^2\theta\,\dd\varphi^2)
\end{equation}
Assuming this metric solves Einstein's equations with the energy-momentum tensor
\begin{equation}
	T^{\mu}_{\;\nu} = \text{diag}(-\epsilon, p_r, p_\theta, p_\phi),
\end{equation}
we aim to derive its rotating counterpart without invoking any complex coordinate transformations.

Following the method proposed in Ref.~\cite{Azreg-Ainou:2014pra}, we promote the metric functions \( G, F, H \) to functions of \( (r,\theta,a) \), denoted \( A, B, \Psi \), such that
\begin{equation}
	\lim_{a \to 0} A = G(r), \quad \lim_{a \to 0} B = F(r), \quad \lim_{a \to 0} \Psi = H(r)
\end{equation}

We define a master function
\begin{equation}
	K(r) \equiv \sqrt{\frac{F(r)}{G(r)}}H(r),
\end{equation}
and rewrite the rotating metric in the Boyer-Lindquist coordinates as
\begin{align}
	\dd s^2 = &-\frac{(F H + a^2\cos^2\theta)\Psi}{(K + a^2\cos^2\theta)^2}\dd t^2 + \frac{\Psi\,\dd r^2}{F H + a^2} + \Psi\,\dd \theta^2  - 2a\sin^2\theta\left[\frac{K - F H}{(K + a^2\cos^2\theta)^2}\right]\Psi\,\dd t\,\dd \phi \nonumber \\
	& + \Psi\sin^2\theta\left[1 + \frac{a^2\sin^2\theta(2K - F H + a^2\cos^2\theta)}{(K + a^2\cos^2\theta)^2}\right] \dd \phi^2
\end{align}

The function \( \Psi(r,\theta,a) \) is determined by ensuring the metric satisfies Einstein’s field equations. Specifically, for an imperfect fluid rotating about the $z$-axis, \( \Psi \) satisfies the following two differential equations
\begin{subequations}
\begin{align}
	&(K+a^2\cos^2\theta)^2 \left[3\left(\partial_{r}\Psi\right)\left(\partial_{\theta}\Psi\right) - 2\Psi \left(\partial_{\theta}\partial_{r}\Psi\right)\right] = 3a^2\left(\partial_{r}K\right)\Psi^2 \\
	&\left[\left(\partial_{r}K\right)^2 + K\left(2 - \partial_{r}^{2}K\right) - a^2\cos^2\theta\left(2 + \partial_{r}K\right)\right]\Psi  + \left(K+a^2\cos^2\theta\right)\left[4\cos^2\theta \left(\partial_{\theta}\Psi\right) - \left(\partial_{r}K\right)\left(\partial_{r}\Psi\right)\right] = 0
\end{align}
\end{subequations}
For static geometries with \( G = F \) and \( H = r^2 \), such as regular black holes, a solution is given by
\begin{equation}
	\Psi = r^2 + a^2 \cos^2\theta
\end{equation}
The resulting rotating metric reduces to a generalized Kerr-like form
\begin{align}
	\dd s^2 &=- \left(1 - \frac{2f(r)}{\rho^2}\right)\dd t^2 + \frac{\rho^2}{\Delta} \dd r^2 + \rho^2 \dd \theta^2 -\frac{4a f(r)\sin^2\theta}{\rho^2} \dd t\,\dd \phi + \frac{\Sigma\sin^2\theta}{\rho^2} \dd \phi^2
\end{align}
where the auxiliary functions are defined as
\begin{subequations}
\begin{align}
	\rho^2 &= r^2 + a^2 \cos^2\theta\\
	2f(r) &= r^2[1 - F(r)] \\
	\Delta &= r^2 F(r) + a^2 \\ 
	\Sigma &= (r^2 + a^2)^2 - a^2\Delta \sin^2\theta
\end{align}
\end{subequations}
To verify the solution satisfies Einstein’s equations, we construct a tetrad basis \( \{e^\mu_t, e^\mu_r, e^\mu_\theta, e^\mu_\phi\} \) adapted to the rotating geometry and compute the Einstein tensor in this frame. The energy-momentum tensor takes the form:
\begin{equation}
	T^{\mu\nu} = \epsilon\, e^\mu_t e^\nu_t + p_r\, e^\mu_r e^\nu_r + p_\theta\, e^\mu_\theta e^\nu_\theta + p_\phi\, e^\mu_\phi e^\nu_\phi,
\end{equation}
where \( e^\mu_t \) is the four-velocity of the rotating fluid. The frame components of the Einstein tensor exactly reproduce this structure, confirming the consistency of the solution with a rotating imperfect fluid source.

Thus, the method provides a general and physically consistent prescription to obtain rotating geometries from static spacetimes while preserving the field equations, causal structure, and energy conditions—without relying on the ambiguities of complexification inherent in the NJA.

\subsection{ Black hole power in terms of BH angular frequency}

\subsubsection{Magnetic flux configuration}
It is a well-established fact that a strongly magnetised relativistic jet does not find it simple to self-collimate. For example, the configuration of the equilibrium field surrounding a spinning BH that is isolated and threaded by a magnetic field (which is generated by external currents) adopts the shape of a split monopole\cite{Blandford:1977ds}. Evidence of self-collimation is evident only at proximity to the polar axis\cite{Tchekhovskoy2009EFFICIENCYOM}.Consequently, to generate a jet that collimates the majority of the energy emitted by the black hole, it is essential to incorporate an external confining medium. The confining agent may be the gas within an accretion disc, a corona, or a wind originating from the disk's inner regions. Since numerical simulations of both the jet and the confining medium is quiet challenging,  the conventional method of implementing a rigid axisymmetric wall with a defined shape, necessitating the jet to remain within the wall, has usually been adopted in literature\cite{Tchekhovskoy2009MagnetohydrodynamicSO}. The configuration of the wall is determined by two parameters: (a) An index $\nu$ that determines the asymptotic shape of poloidal field lines. This parameter varies from $\nu=0$, indicative of a monopole field shape, to $\nu=1$, representative of a paraboloidal jet. In a practical system, $\nu$ would be determined by the radial pressure profile of the confining medium. (b)A transition radius $r_{0}$ is established so that for $r>r_{0}$, the jet is monopolar, and for $r<r_{0}$, it adopts the shape dictated by $\nu$. The parameter $r_{0}$ enables the examination of scenarios where confinement is effective solely beyond a specific distance from the black hole. Given these two parameters, the wall exhibits the following configuration in polar $(r,\theta)$ coordinates inside the Boyer-Lindquist framework:
\begin{flalign}
1-\cos\theta=\left(\dfrac{r+r_{0}}{r_{H}+r_{0}} \right)^{-\nu}
\end{flalign}
where $r_{H}$ is the event horizon of the black hole deterimined by solving the equation $\Delta(r_{H})=0$ (for more than one solution the greatest positive solution is considered). For $r\ll r_{0}$, which implies $(1-\cos\theta)\ll 1$, suggests $\theta=\dfrac{\pi}{2}$, i.e., the wall is positioned along the equatorial plane, similar to a split monopole. For $r\gg (r_{H},r_{0})$, $\theta\propto r^{-\nu/2}$, which is associated to a generalised paraboloid. In all these models, the wall intersects the horizon at the equator. This implies that a razor-thin disc, which subtends zero solid angle at the black hole, has been taken into account. For such disks, the \emph{`disk thickness'} $\left(\dfrac{H}{R}\right)=0$.

\begin{figure}[htbp!]
	\centering
	\includegraphics*[scale=0.6]{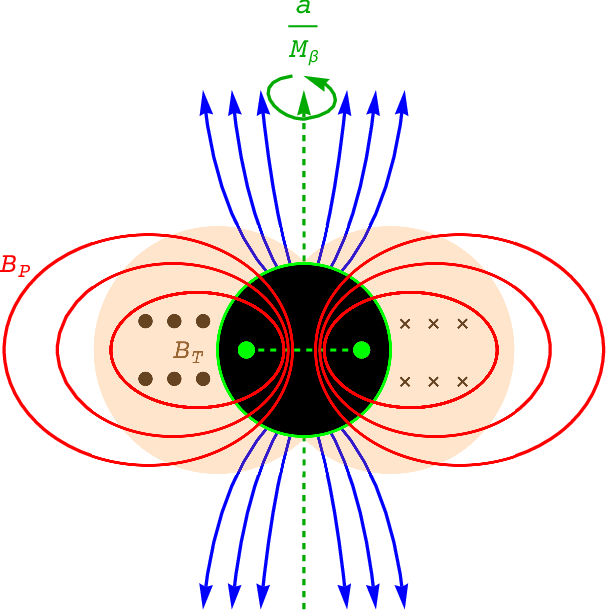}
	\caption{Geometric representation of the poloidal $B_{P}$ and toroidal $B_{T}$, magnetic fields superimposed on the horizon + ergosphere structure.}\label{Fig:TPF}
\end{figure}

 After selecting the  geometry of the wall, we designate the poloidal flux function of the initial magnetic field configuration as
\begin{flalign}\label{Eq:FF1}
\Psi=\left(\dfrac{r+r_{0}}{r_{\rm H}+r_{0}} \right)^{\nu}\left(1-\cos\theta \right) 
\end{flalign}
Here, $\Psi$ is conserved along each field line, and $\Phi=2\pi \Psi(r,\theta)$ is the poloidal magnetic flux through a toroidal ring at $(r,\theta)$. \ref{Eq:FF1} inherently represents a total flux of $\Phi_{\rm tot}=2\pi$ in the jet. In this model, $\Phi_{\rm tot}$ is independent of spin, so the quantity of magnetic flux passing through the black hole remains constant across varying spins. The outermost field line, indicated by $\Psi=1$, conforms to the contour of the wall. This specific field line is consistently situated at the boundary. One need to know that the initial magnetic field lacks a toroidal component\cite{Tchekhovskoy2009BLACKHS}. The representation of magnetic field threading and structure of poloidal and toroidal magnetic field have been shown in \ref{Fig:TPF}.

No exact solution are known for the magnetosphere of a rotating black hole. Nevertheless, the poloidal field configurations shown in \ref{Eq:FF1} closely approximate the actual solutions, resulting in a relatively gentle initial relaxation at the commencement of any simulation. Only two linearly independent analytic solutions have been derived for a non-spinning black hole: one corresponds to a monopolar field geometry, represented by \ref{Eq:FF1} with $\nu=0$, while the other is the solution for a paraboloidal field \cite{Blandford:1977ds},
\begin{flalign}
\Psi=\dfrac{\left( \dfrac{r}{r_{\rm H}}-1\right)(1-\cos\theta)-(1+\cos\theta)\ln(1+\cos\theta)}{2\ln 2}	+1 .
\end{flalign}	

The split-monopole solution is applicable to the entirety of the space outside the horizon, whereas the paraboloidal solution is limited to the field lines connected to the black hole.

\subsubsection{Force-free magnetosphere}\label{App_2_2_0}

{
This section provides a concise derivation of the Blandford-Znajeck effect\cite{1977MNRAS.179..433B}.
The original BZ model features a Kerr black hole encircled by a stationary, axisymmetric, force-free, magnetic plasma. 
To describe the force-free magnetosphere along with the Maxwell's equations with a source, the electromagnetic field should satisfy the following  conditions\cite{Gralla:2014yja,10.1093/mnras/167.3.457}
\begin{subequations}
\begin{flalign}
	\nabla_{\alpha}F^{\alpha\beta}=J^{\beta}\\
	F_{\alpha\beta}J^{\beta}=0\\
	\mathcal{G}_{\alpha\beta}F^{\alpha\beta}=0\\
	F_{\alpha\beta}F^{\alpha\beta}>0
\end{flalign}
\end{subequations}
whre, $J^{\beta}$ is the four-current, $\mathcal{G}^{\alpha\beta}$ is the dual tensor of $F_{\alpha\beta}$.

In the force-free approximation, the energy-momentum tensor pertains solely to the electromagnetic field, disregarding matter. Hence the field energy-momentum tensor can be given by
\begin{flalign}
	T^{\mu\nu}=T^{\mu\nu}_{EM}={F}^{\mu\rho}\tensor[]{{F}}{^{\nu}_{\rho}}-\dfrac{1}{4}g^{\mu\nu}{F}^{\sigma \rho}{F}_{\sigma \rho}\label{Eq:TMN}
\end{flalign}
where $F_{\mu\nu}=\nabla_{\mu}A_{\nu}-\nabla_{\nu}A_{\mu}$ is the Faraday tensor and $A_{\mu}$ is the vector potential. The equation of motion is given by 
\begin{flalign}
	\nabla_{\mu}T^{\mu\nu}_{EM}=0\label{Eq:TMN_S}
\end{flalign}
Considering the force-free condition and assuming that $A_{\mu}$ is independent of the coordinates $t$ and $\phi$, the Faraday tensor can be written as 
\begin{flalign}
	F_{\mu\nu}=\sqrt{-g}\begin{bmatrix}
		0& -\omega B^{\theta} & \omega B^{r} & 0\\
		\omega B^{\theta} & 0 & B^{\phi} & - B^{\theta}\\
		-\omega B^{r} & - B^{\phi} & 0 & B^{r}\\
		0& B^{\theta} & -B^{r} & 0
	\end{bmatrix}\label{Eq:EM}
\end{flalign}
where
\begin{flalign}
	\omega(r,\theta)=-\dfrac{\partial_{\theta}A_{t}}{\partial_{\theta}A_{\phi}}=-\dfrac{\partial_{r}A_{t}}{\partial_{r}A_{\phi}}
\end{flalign}
is the angular frequency of the electromagnetic field lines or known as {\it electromagnetic angular velocity}. $B^{i}\equiv \mathcal{G}^{i0}$ is the magnetic field, $\mathcal{G}^{\mu\nu}$ is the dual tensor of the Faraday tensor.

Then the conserved electromagnetic energy and angular
momentum flux are written as
\begin{subequations}
\begin{flalign}
	\mathcal{E}^{\nu}\equiv -\tensor[]{T}{^\nu_{t'}}\\ 
	\mathcal{L}\equiv \tensor[]{T}{^{\nu}_{\phi'}}
\end{flalign}	
\end{subequations}

Using \ref{Eq:GKS} for spacetime metric and \ref{Eq:EM} for electromagnetic field, from \ref{Eq:TMN} , we have
\begin{flalign}
	\mathcal{E}^{r}=-\omega(B^{r})^{2}\left\{\omega\left[r^{2}+a^{2}-\Delta \right]-a\right\}\sin^{2}\theta - \omega \Delta B^{r}B^{\phi}\sin^{2}\theta
\end{flalign}
At event horizon, we have $\Delta|_{r=r_{H}}=0$. So at event horizon, we obtain
\begin{flalign}
	\mathcal{E}^{r}|_{r=r_{H}}=-\omega(B^{r})^{2}\left\{\omega\left(r_{H}^{2}+a^{2} \right)-a\right\}\sin^{2}\theta 
\end{flalign}
The energy flux from the black hole is defined by
\begin{flalign}
	\mathcal{F}_{E}(\theta)=\mathcal{E}^{r}(r_{H},\theta)
\end{flalign}
The total rate of energy extraction from the black hole
is given by the integral over the event horizon surface
\begin{flalign}\label{Eq:PBZ}
	P_{\rm BZ}=\iint \sqrt{-g}\mathcal{F}_{E}(\theta) \dd\theta \dd\phi'=4\pi \int_{0}^{\pi/2}\sqrt{-g}\mathcal{F}_{E}(\theta)\dd\theta 
\end{flalign}
In order to evaluate the integral in \ref{Eq:PBZ}, we need to solve \ref{Eq:TMN_S} for determining $B^{r}$ and $\omega$. However, this determination of the solution is quiet non-trivial. The conventional method involves deriving an exact solution of \ref{Eq:TMN_S} for the Schwarzschild spacetime, followed by a perturbative expansion in terms of $a$ or $\Omega_{\rm H}$ to obtain the rotating solution.

}

\subsubsection{Power output}\label{App_2_2}
 We ascertain the power output of a rotating black hole situated within an externally applied split-monopolar magnetic field. The magnetic field is defined by the flux function \ref{Eq:FF1} with $\nu=0$ and $r_{0}=0$.  The power output density assessed near the horizon of a rotating black hole is
\begin{flalign}\label{Eq:Out}
\mathfrak{F}_{E}(\theta)&=\mathcal{E}^{r}|_{r=r_{\rm H}}=-\omega(B^{r})^{2}\left[ \omega(r_{H}^{2}+a^{2})-a\right]\sin^{2}\theta=\omega (B^{r})^{2}a\left(1-\dfrac{\omega}{\Omega_{H}} \right)\sin^{2}\theta\nonumber\\
&=(B^{r})^{2}\omega\left( \Omega_{H} -\omega\right)(r_{H}^{2}+a^{2})\sin^{2}\theta
\end{flalign}	
where $\omega$ is the angular frequency of the magnetic field lines at the black hole horizon, $\Omega_{H}$ is the black hole spin-frequency at the event horizon and $B^{r}$ is the radial field strength at the hoprizon.

To ascertain the power output \ref{Eq:Out}, it is essential to identify two parameters as functions of polar angle at the black hole horizon: the radial magnetic field, $B^{r}$, and the field line angular frequency $\omega$. The differential relationship between the magnetic flux element $\dd\Phi$ through the black hole horizon and the radial magnetic field at the horizon is as follows
\begin{flalign}\label{Eq:Out1}
\dd \Phi=2\pi \dd \Psi=2\pi B^{r} \sqrt{-g}~\dd \theta
\end{flalign}
where $g$ is the determinant of the rotating spacetime metric. Assuming that the deviations from a perfect split-monopole in the magnetic field are of higher order, we disregard them and, by applying the difference to the flux function shown in \ref{Eq:FF1} (with $\nu=0$ and $r_{0}=0$) as per \ref{Eq:Out1}, we derive a result equal to that in flat space
\begin{flalign}\label{Eq:Br}
B^{r}=\dfrac{\Psi_{\rm tot}}{r^{2}}
\end{flalign}
where we disregarded terms of order $\Omega_{H}^{2}$ and above. The distribution of $\omega$ must be self-consistently established by solving the non-linear equations that describe the equilibrium of electromagnetic fields (this is accomplished numerically in \cite{Tchekhovskoy2009BLACKHS}); however, we will provide a straightforward yet precise approximation based on the energy argument. Assume the system selects a distribution of $\omega$ that results in an extremum in black hole  power output in \ref{Eq:Out}. This value is evidently
\begin{flalign}\label{Eq:OH}
	\omega=\dfrac{1}{2}\Omega_{H}
\end{flalign}
as it maximizes the black hole power output\cite{2000NCimB.115..795B}. Based on the results of the numerical simulations\cite{Tchekhovskoy2009BLACKHS}, this estimate is extremely near to the actual solution for the split-monopolar geometry, despite the fact that it is rather straightforward. Inserting \ref{Eq:Br} and \ref{Eq:OH} into the power  output density \ref{Eq:Out}, we obtain
\begin{flalign}\label{Eq:PP}
	\mathfrak{F}_{E}(\theta)=\left(\dfrac{\Omega_{H}}{2}\right)^{2}\left(\dfrac{\Psi_{\rm tot}}{r^{2}} \right)^{2}\times 2f(r_{H})\sin^{2}\theta
\end{flalign}
By integrating the power output density in angle in the same manner that we integrated the magnetic field in \ref{Eq:Out1}, we are able to acquire the whole power output into jets that have an opening angle of $\theta_{j}$:
\begin{flalign}\label{Eq:Pout}
P=2\times 2\pi\int_{0}^{\theta_{j}}\mathfrak{F}_{E}(\theta)\sqrt{-g}|_{r=r_{H}}\dd\theta
\end{flalign}
where the additional factor of two is responsible for the fact that there are two jets, one present in the northern hemisphere and one present in the southern hemisphere. It is important to take note that we are curious about the possibility of expanding power up to the second order in $\Omega_{H}$. Given that the factor $\mathfrak{F}_{E}(\theta)\propto \Omega_{H}^{2}$ is already of the second order in $\Omega_{H}$, we are able to assess the formula \ref{Eq:Pout} at $r=r_{H}(a=0)$ without compromising the correctness of the calculation. In this case, we can substitute $\sqrt{-g}$ with $r_{H}^{2}\sin\theta$. Following the process of plugging into \ref{Eq:Pout} for $\mathcal{F}_{E}(\theta)$ using \ref{Eq:PP} and evaluating the integral out to $\theta_{j}=\dfrac{\pi}{2}$, which is the computation of the whole power output of the black hole, we obtain the following results
\begin{flalign}
P&=\pi\Psi_{\rm tot}^{2}\Omega_{H}^{2}\times\dfrac{2f(r_{H})}{r_{H}^{2}} \times\int_{0}^{\pi/2}\sin^{3}\theta\dd\theta=\dfrac{2\pi}{3}\left[1-F(r_{H}) \right]\Psi_{\rm tot}^{2}\Omega_{H}^{2}
\end{flalign}
which is accurate upto second order in $\Omega_{H}$. Recalling the magnetic flux $\Phi_{\rm tot}=2\pi \Psi_{\rm tot}$, we obtain
\begin{flalign}\label{Eq:JetPow}
P=k\xi\Phi_{\rm tot}^{2}\Omega_{\rm H}^{2} +\mathcal{O}(\Omega_{H}^{4})
\end{flalign}
where where $\xi\equiv (1-F(r_{H}))$ is the {\it regularization factor} and  $k=1/6\pi$. For paraboloidal scenario, the expression will be same with constant $k=0.044$. Interestingly, this regularization factor appears only for regular black holes. If one include higher order terms, then \ref{Eq:JetPow} becomes
\begin{flalign}
	P_{\rm BZ}=k\xi \Phi_{\rm tot}^{2}\Omega_{\rm H}^{2}~ h\left(\Omega_{\rm H}^{2} \right) 
\end{flalign}
for some function
\begin{flalign}
	h \left( \Omega_{H}^{2}\right)=1+c_{1}\Omega_{H}^{2}+c_{2}\Omega_{H}^{4}+\cdots=\displaystyle{\sum_{i=0}}c_{i}\Omega_{\rm H}^{2i}
\end{flalign}

where $\{c_{i}\}$ are numerical coefficients.

\bibliographystyle{./utphys1}
\bibliography{references}

\end{document}